Jagiellonian University
Cracow
Faculty of Physics, Astronomy
and Applied Computer Science

# Software for physics of tau lepton decay in LHC experiments

Tomasz Przedziński

## Abstract

Software development in high energy physics experiments offers unique experience with rapidly changing environment and variety of different standards and frameworks that software must be adapted to. As such, regular methods of software development are hard to use as they do not take into account how greatly some of these changes influence the whole structure. The following thesis summarizes development of `TAUOLA C++ Interface` introducing $\tau$ decays to new event record standard. Documentation of the program is already published. That is why it is not recalled here again. We focus on the development cycle and methodology used in the project, starting from the definition of the expectations through planning and designing the abstract model and concluding with the implementation. In the last part of the paper we present installation of the software within different experiments surrounding Large Hadron Collider and the problems that emerged during this process.

* *This work is partially supported by EU Marie Curie Research Training Network grant under the contract No. MRTN-CT-2006-0355505, Polish Government grant N202 06434 (2008-2011) and EU-RTN Programme: Contract No. MRTN-CT-2006-035482, 'Flavianet'.*

# *Acknowledgements*

*I would like to thank my advisor and the head of the development team, prof. Zbigniew Wąs, who gave me a chance to take part in this project. This thesis and the project itself would not be possible without his patience and guidance.*

*I am grateful Elżbieta Richter-Wąs for introducing me to high energy physics and for her help during last three years. I would also like to thank Nadia Davidson and Gizo Nanava for their cooperation in project development.*

*It would not be possible to finish the project without the help of Piotr Golonka and all devoted users, in particular Anna Kaczmarska,* Dmitri Konstantinov, *Sami Lehti, Eric Torrence, Marcin Wolter Julia Yarba and* Oleg Zenin*.*

Contents

# 1. Overview

## *1.1* Introduction

High energy physics offer excellent environment for understanding development of large software engineering projects. The development process changes rapidly as even the team responsible for the project may completely change during software development. Consequently, the development time becomes inappropriately large in comparison to the short lifetime of the software version use, creating problems that even agile software development methodologies cannot resolve. At the same time, history of the projects is usually large and often extends over decades. The style of the code differs from one part to another and a large amount of testing procedures, as well as old versions of the algorithms, still remain in the code, commenting not only the algorithm itself but its history of changes and prototypes for future as well. That is why adapting to new technology or rewriting the code to new programming language becomes a hard task. The tests needs to be redone and new members of the team need to learn all the basics regarding the project. Some parts of the original comments must remain intact, as they might be needed in future. It requires a lot of work, which invalidates previously written documentation and leads to loss of experience invested so far. That is why such decisions should be made only when it is absolutely necessary for the survival of the project.

The project management also differs from commercial applications, as there is no central management regarding its content or financial aspect. It is a challenge for community exceeding sometimes even thousand people. The crucial element in development of project spanning many years of work is training new members and wise distribution of their acquired knowledge.

My work, as a trainee in such team, focused on the project regarding physics aspect of $\tau$ lepton production and decay as well as creation of tests of Monte Carlo generators. I often had to follow other people designs thus my understanding of the work style became clear later in the project. In fact, it became more apparent during my work on the thesis itself. That is also a reason why some parts of this project I do not fully understand. For example, I have not gained experience in software transformation related to evolution of physics model and problems related to it. It comes from the lack of experience and the fact that I was participating in only a step of the whole project development. In result the reason for decisions regarding the relation between `FORTRAN` code or the electroweak corrections module and the C++ interface were, for a longer time, not clear to me. Nonetheless, I blindly had to follow the design only to realize motivation behind these decisions later, as my understanding of the project background and the uniqueness of environment for which the project was intended expanded.

This thesis summarizes effort documented in reference [1], describing development of a module implementing $\tau$ decays into a simulation chain. It is also an attempt to explain software development strategies used in this project and to point out similarities between methodology used during development cycle and agile approach.

Reference [1] represents the fundaments of this thesis and documents the results of work of our team on `TAUOLA` C++ interface. Reference [2] describes a new version of `MC-TESTER`, which was a crucial testing tool used in creation of this project. Reference [3] describes the possible future extensions to this work. I am a co-author of these three references. The con-tent of [2] and [3] will not be described here. I will not present documentation of the program I was developing, it is documented in reference [1]. My thesis will be focused on general scheme applied in the work. In this respect it can be understood as continuation of effort by P. Golonka [4] to systematize the approach in program development. An outline of the PHOTOS C++ Interface presented in [5] regards the next activity in High energy physics in which I will be taking part. Both MC−TESTER and PHOTOS has already been subject of extensive studies [4 pp. 149-167, 193-205] and will not be the subject of this paper. The base of the project, TAUOLA FORTRAN, is well-known, fully developed software used by a range of specialists. This work will be focused on the planning and development strategy used to create TAUOLA C++ interface introducing new C++ event record, new functionality and new physics process to generation sequence of τ decay. Let me follow with a citation from an introduction to reference [1].

## *1.2* The black-box approach

In the present day experiments at High energy physics accelerators, interpretation of results becomes increasingly involved. Not only the detector response is complex, but also some theoretical effects need to be removed. Otherwise, results are difficult to interpret for the non-specialist. For that purpose the concept of work with realistic and idealized observables was introduced as well as finally with pseudo-observables which can be easily understood by theorists, such as W, Z masses or couplings. Good examples of such approaches were measurements of the two-fermion final states at LEP. Because of increasing precision of the experimental measurements, definitions of quantities to be measured, simple at first, later evolved into several options [6], each based on the properties of individual detectors and each requiring individual discussion of the systematic error. One could assume that if all theoretical effects are embodied into one theoretical black-box and, experiments while using it tune parameters (representing pseudo-observables) to the data, interpretation of the observed effects could be separated into theoretical and experimental components. Unfortunately, this strategy is limited, as it leaves little room for cases where theory and experimental effects are convoluted: size and even nature of the theoretical corrections depend on the experimental conditions. Such discussion on observables involving τ decays can be found in [7]. Recently, discussion for the physics of τ production and decay at low energies where similar aspects are addressed, was presented in [3].

For LHC experiments, τ decays themselves are not of primary interest, but rather will be used to measure properties of τ production processes. Physics effects necessary for the prediction of hard processes at the LHC experiments can be separated into several parts, among them: parton showers, the underlying event and structure functions, final state QED bremsstrahlung, QED bremsstrahlung interference between initial and final states and finally the hard process including electroweak corrections. Such separation is not only for the convenience of organizing theoretical work but also provides efficient and flexible component in the framework for experimental data analysis strategy (see e.g. [8]). Some of such building blocks are of genuine theoretical interests; some others are not so much. The hard process usually depends on the parameters intended for the measurement, e.g. W or Higgs mass or new coupling constants. Other building blocks may be less interesting; nonetheless, they may affect the results of measurements. This is certainly true in the case of the underlying event or missing transverse energy or $p_T$ distributions generated from parton showers [6]. It may also be the case for QED final state bremsstrahlung or initial-final state interference, where potential difficulties may be expected [9] and predictions may need to be fixed with the help of experimental data.

The black-box approach, where all simulation segments are put together by theorists, may look advantageous to the experimental user. However, in such a case one has less flexibility to distinguish experimental effect from the theoretical ones, thus limiting control on the systematic errors. The particular problems may be left unnoticed. Typically, the difficulties will not affect all observables. Unfortunately, complications tend to show up only when more detailed discussion on the systematic errors of experimental analysis is performed.

## *1.3* Main goals

The most important goal of this project is to introduce the software for τ decays to different experiments environments while still allowing low energy physicists to modify the core of the project. In order to do it, the main engineering goal is to adopt existing `FORTRAN` code to new C++ even records, so that it can be attached to any Monte-Carlo program where τ's are generated. The project will focus on adapting to `HepMC` event record [10], as it seems that `HepMC` will remain a generally accepted standard for data structure used in phenomenology of high energy physics for the near future. `HepMC` introduces new data structure as well as new exceptions that need to be handled. Thanks to new event records, introducing τ decays to simulation will become more flexible than its `FORTRAN` predecessor. It will allow extending generation procedure with electroweak corrections and algorithms for handling of weighted events.

Since the slightest change in the code may influence numerical results of the existing Monte Carlo generator, it is important to organize a good testing environment. A wide range of tests will be needed in order to validate that the communication with the `FORTRAN` code is correct as well as the output produced in different event record is acceptable. Since the base of the tests for `FORTRAN TAUOLA` already exists, these tests will have to be moved to C++. Moreover, for each new feature introduced to the original generation process a new set of tests will be needed as well. Using new environment and new event record a wide range of new tests can be performed for further validation of τ decays[1].

The most challenging task will be to coordinate the work between people of different fields of expertise. The development will require cooperation with people from different countries and living in different time zones, which must be included during the planning. As soon as the first version of the code is present, software will need to be set up under several testing environments of different experiments, therefore a user friendly, flexible installation procedure will be needed. We have decided to use autoconf [11] as a mean to create configuration procedure for wide range of platforms. Automake scripts has been provided as an alternative method of installation.

Finally, both the code and the physics content of the software require extensive documentation and the webpage used for promotion. This will be achieved using doxygen [12] software allowing automatic generation of source code documentation and introducing an easy way to build a front page of the website. Thanks to this, after setting up an SVN repository, we can automatically create a new release archive that can be distributed to pilot users.

---

[1] These tests may represent starting point for some physics analysis where detector effects are taken into account, see e.g. [37].

# 2. Starting point

## *2.1* Software in high energy physics

The LHC Computing Grid (LCG) software database stores hundreds of projects, millions lines of code. It is designed to help distribute new versions of code for projects, like ours, within experiments surrounding LHC. Internally, collaborations have their own way of supporting external packages. Integrating a new program designed to cooperate with several already existing projects introduces a list of dependencies that have to be taken into account and each of them can create a potential problem. The logical structure of these projects can sometimes be very hard to understand and in most cases should not be studied at all. One of the tasks is to decide which parts of the vast amount of knowledge considering these projects are actually relevant to the project and need to be learned, which in most cases is very hard to find out.

The target users for the software are physicists interested in analysis of $\tau$ production. This element has to be taken into account when writing the code of the project, as the code directed to physicists is different from the regular, commercial code. In most cases, there is no need for the user to modify any aspect of the code, as the user is more interested in the complete product itself. In case of High energy physics the approach is exactly opposite than the one expected. The code should be written in such way that the physicists with a basic programming knowledge should be able to understand the crucial parts of the algorithms and modify it if needed. That is why most of the projects contain a lot of obsolete code or even hundreds of lines commented out to serve as information on how the project evolved or how different solutions can be applied. Such comments may in future prove to be very useful for modifying or debugging the code.

That being said, even if better methods of writing the code structure or the algorithm exists, the clarity of the code and flexible modification of its parts gets the priority over the cleaning and optimization. This concern the basic data structure as well since the most optimal method of storing the information might not be as clear as the one that is closest to physical model. The aim of designing such data structure is to balance the performance of the code with its ease of modification, emphasizing the latter rather than the former aspect. However, in most cases, this compromise is hard to achieve and parts of the code remain in disarray, which might confuse future reader. Having read a lot of source code from such projects, I have realized that rough, puzzling code is one of the aspects of these projects that cannot be avoided. Even though it was hard to accept at first, I have learned that in case of High energy physics software, the best solution is not always the right one, but more importantly – that the code that is easily readable by the programmer is not easily read by a physicist and vice versa.

Apart from that, the development of the software is mixed with the progress of the testing environment build throughout the project development as well as physics model itself. Since the basic model is often based on insufficient assumptions, new aspects must be introduced as soon as part of the model changes which greatly influence the development cycle. Considering that such projects have a limited number of testers, it is very hard to detect bugs introduced by such frequent changes.

All of the problems presented above affect both planning and development of the project and are the reasons why the strategy used in this project differs from any known methodology.

## 2.2 Description of FORTRAN TAUOLA

For the purpose of generation of τ decays themselves, the TAUOLA library, as described in [13; 14; 15] is used. This part of the code is expected to be a black-box for the High Energy experimental user. At present, from the technical side, the black-box consists of the same FORTRAN code as described in [16]. Such organization makes it easy for low energy phenomenologists or experimentalists to continue their work on this part of the code, such as the activities proposed by us in [3], leading to the new parameterization of hadronic τ decay currents becoming available for High Energy experimental users in a rather straightforward way.

There is virtually no gain in rewriting the FORTRAN TAUOLA to C++, while on the other side such activity could lead to number of errors and would require a large amount of time dedicated to testing and documenting newly introduced changes. Keeping the code in FORTRAN allows the base of users of previous versions of TAUOLA to be able to modify appropriate aspects of the generator while introducing new functionality through the interface to C++ event records. Since we expect FORTRAN part of the code to be modified by low energy physicists, the decision was made to leave TAUOLA FORTRAN intact especially that this arrangement does not pose any troubles for development of the interface.

The interface will overwrite `filhep_` method for event record I/O, changing the event record used by TAUOLA FORTRAN and use `decay_` method for generating τ decays. It will also need several common blocks along with their accessors for initialization and setup. The summary of all FORTRAN methods used in the project is located in Appendix of reference [1].

## 2.3 Applying electroweak corrections

Physics of spin correlations in τ pair productions is defined since long. Principles of the necessary algorithms are given in reference [13]. Nothing new is necessary for today applications. However, density matrix for the produced τ pair must be provided. In our interface example density matrices for basic processes of interest at LHC are given. This is in most cases sufficient, however in case of τ pair production through intermediate Z and virtual gamma born level predictions are not sufficient. This is known since KKMC [17] or even KORALZ [18] times, as in these cases one needs a method to calculate τ pair density matrix including electroweak corrections.

Fortunately, specialized libraries like SANC [19] exist. With their help density matrix as a function of incoming quark or lepton flavour center of mass energy and angle can be calculated. Such algorithms have already been created in FORTRAN and extensively used by KORALZ and KKMC. Using experience from these projects, a separate module will be created introducing electroweak corrections to the interface.

## 2.4 HepMC event record

As stated in reference [1], in adapting both TAUOLA and MC−TESTER to the C++ event record format the difference between the HEPEVT event record used in the FORTRAN ver-sion of TAUOLA interface and HepMC event record that is used for the C++ based interface has to be taken into account. In the first case, the whole event was represented by a common block containing a list of particles with their properties and with integer variables denoting pointers to their origins and descendants. The HepMC event structure is built from vertices, each of them having pointers to their origins and descendants. Links between vertices represent particles or fields. In both, FORTRAN and C++ cases, the event is structured as a tree[2] the necessary algorithms are analogous, but nonetheless different.

In HepMC, an event is represented by a GenEvent object, which contains all information regarding itself, including event id, units used for dimensional quantities in the event and the list of produced particles. The particles themselves are grouped into GenVertex objects allowing access to mother and daughter particles of a single decay. Vertices provide an easy way to point to the whole branch in a decay tree that needs to be accessed, modified or deleted if needed. The information of a particle itself is stored in a GenParticle object containing the particle id, status and momentum as well as information needed to locate its position in the decay tree. This approach allows traversing the event record structure in several different ways.

In general, event record represents a tree structure, however due to physics constraints there are options that break this rule. In case of theoretical simulation, when writing to event record, some Monte Carlo generators include extra information regarding the decay process. While normally such information might be helpful to determine what exactly happened during event generation, the additional information might be misleading requiring the algorithms to account for such generator behavior. In case of experimental data, the event structure must be reconstructed using data taken from different types of detectors. The output might not always represent a consistent structure, as there is always a possibility of an error during reconstruction step. Such special cases of event record structures require extensive attention. From a point of view of a software engineer, some of them might look similar to numerical error or a faulty algorithm routine, which is why such errors should be analyzed both from physics and from algorithmic point of view.

The HepMC event record format is evolving with time, making it necessary to adapt the code to the new versions. The project started with HepMC version 2.03; however, the present version 2.05 already introduced several differences to how the event record is stored, such as the different standards of units used to represent particle momentum or vertex position in space. This example further emphasizes on need to divide the event record from the physics content of the software.

Seeing as different standards are quickly becoming more specialized, we have also envisaged the possibility that HepMC may one day be replaced by another standard of event record, and we have provided an easy way to extend the interface to a possible new event record standard.

For this purpose, an abstract interface will be created containing full functionality of $\tau$ decays. It will be then extended for HepMC event record and for any other event record that might be needed in future. The advantage writing in an object oriented programming style allows us to create a modular solution handling any possible future exceptions.

---

[2] At least in principle, because in practice its properties may be rather of the graph of not universally defined properties. This makes our task challenging.,

## 2.5 Event record standard

Evolution of the HepMC format itself is not a crucial problem. In contrary, conventions how physics information is filled into HepMC represent the source of main technical and physics challenge for our interface.

While many Monte Carlo generators (e.g. PYTHIA 8.1[20], HERWIG++ [21]) store events in HepMC format, the representations of these events are not subject to strict standards, which can therefore vary between Monte Carlo generators or even physics processes. Some examples of these variations include the conventions of status codes, the way documentary information on the event is added, the direction of pointers at a vertex and the conservation (or lack of conservation) of energy-momentum at a vertex.

Below is a list of properties for basic scenario we have observed in Monte Carlo generators used for testing the code. This list serves as a declaration for convention of HepMC filling, which the interface should be able to interpret correctly.

- 4-momentum conservation is assumed for all vertices in the event record.
- Status codes: only information whether given particle is incoming, outgoing or intermediate will be used,
- Pointers at a vertex are assumed bi-directional. That is, it is possible to traverse the record structure from mother to daughter and from daughter to mother along the same path.

Detailed conventions for the actual filling of physics information into HepMC format is defined by authors of each Monte Carlo program. Usually they modify some of the above properties.

In future, an important special case of event records filling with information extracted from experimentally observed event (e.g. $Z \to \mu^+ \mu^-$ modified later to $Z \to \tau^+\tau^-$) should be allowed. Obviously, a new type (or types) of HepMC filling will then appear.

## 2.6 Testing routine

For testing purposes, we have focused our attention on PYTHIA 8.1 universal Monte Carlo generator. As written in introduction to its documentation[20]: *The PYTHIA program is a standard tool for the generation of high-energy collisions, comprising a coherent set of physics models for the evolution from a few-body hard processes to a complex multihadronic final state. It contains a library of hard processes and models for initial and final state parton showers, multiple parton-parton interactions, beam remnants, string fragmentation and particle decays. It also has a set of utilities and interfaces to external programs.* In particular, PYTHIA 8.1 can produce output in HepMC event record format. The simulation routine used by PYTHIA 8.1 consists of few basic steps:

1) Choose simulated process (proton-proton collision, e+e- collision, ...)
2) Set center of mass energy (1GeV, 7TeV, 14TeV, ...)
3) Select generation mode / intermediate boson (W, W+, Z0, H, H+, ...)
4) Choose an output event record (HEPEVT, HepMC, ...)
5) Generate large set of events
6) Analyze particles of interest

The PYTHIA configuration system allows turning off internal computation of τ decays. This way, having a complete event record without the decayed τ leptons, we can attach our software, produce decay based on data from event record, and save the output inside the same event. The generation procedure of single τ decay would look like this:

a) Get all information from event record needed for the decay
b) Generate τ
c) Modify its kinematic according to its siblings, mothers and grandmothers
d) Decay τ
e) Compute daughters kinematic
f) Decay any unstable daughters (if needed)
g) Apply post-generation effects (photon emission, ...)

After a large set of data is generated, we can then validate them by analyzing the decays generated by our software and compare them with verified results. For this purpose, MC−TESTER, described in the next section, can be easily attached to testing routine. The relations between each component are presented on *Fig. 1*.

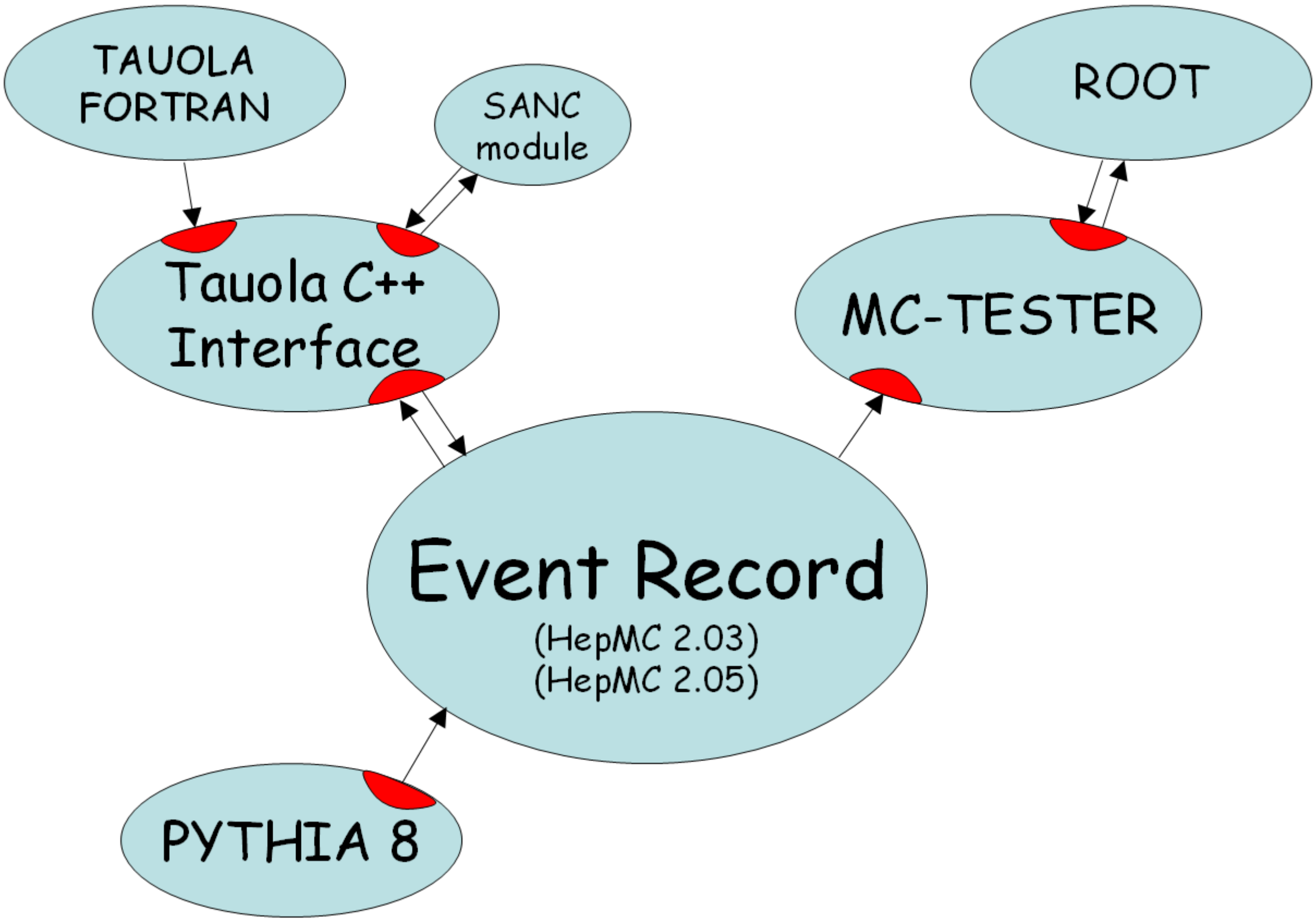

*Fig. 1 Diagram of relations of different components. By design, event record serves as the means of communication between each module in the analysis sequence. A red part of a component symbolizes a separate module or interface used for communication while arrows represent information flow.*

Following the previously described PYTHIA 8.1 generation routine, we can easily prepare an example of use of the interface as well as basic environment for testing purposes, by using the output event record created in step 4 as an input of our interface and by using MC−TESTER for analysis (step 6). As a result, a complete analysis process would look like this:

1) Define all parameters for PYTHIA, TAUOLA interface and MC-TESTER
2) Suppress τ decays generated by PYTHIA.
3) Initialize all three modules
4) For each event to be analyzed:
   a) Generate new event using PYTHIA
   b) Save generated event to HepMC
   c) Invoke TAUOLA interface to apply τ decays
   d) Analyze event record using MC-TESTER
5) Finalize analysis procedure
6) Save analysis results to ROOT file.

### 2.7 MC-TESTER

The huge amount of distinct distributions produced for analysis purpose would be impossible to interpret without a proper tool. Focusing on τ decays alone, we have a huge set of events and a set of distributions that need to be compared. For this purpose, we have prepared MC-TESTER [2], which performs such comparison in a semi-automatic way.

After generation of each event by a given Monte Carlo system, the content is searched by MC-TESTER for the decay of the particle to be studied. Once it is found the appropriate data is collected and stored in the form of automatically generated histograms and tables. A booklet of comparisons is created: for every decay channel, all histograms present in the two outputs are plotted and parameter quantifying shape difference is calculated. Its maximum over every decay channel is printed in the summary table.

Due to flexibility of this software, it is easy to introduce specialized tests for specific physics effect such as electroweak corrections or transverse spin correlation. For the purpose of benchmarking our projects, we had to design and maintain tests. Some of those tests gradually evolved into the new version of MC-TESTER. The wide range of uses of MC-TESTER for testing our project was the main reason for the decision to develop a new version of this tool as a part of the process of creating the main project. The full development cycle, has already been finished and is documented in [2] therefore, it will not be described in this paper, even though I am co-author of its present version.

At present MC-TESTER uses ROOT software analysis framework [22] and is designed to be easily extensible by a means of ROOT scripts. Its functionality is nonetheless to a large degree independent. A wide range of options and an easy method to create different tests made MC-TESTER a perfect tool for testing of our software. Because of the tests made during development of the project, new functionality has been added to MC-TESTER and the software has been modified to work with HepMC event record as well. These tests are the fundamental achievement of the new version of MC-TESTER.

For the most basic tests of our software, we have created benchmark files using TAUOLA FORTRAN code to verify that transition from HEPEVT to HepMC event record does not create any unforeseen changes in numerical results. Afterwards, these benchmark files can be used as a base for testing new options introduced in C++ code. The testing process always take large amount of time, however in case of physics software it is an essential part which has to be done from a variety of different perspectives as some significant errors might be very hard to find from regular point of view. MC-TESTER allowed us to cover several different views and to pinpoint such errors in early stages of software development.

## 3. Abstract algorithm

Once starting point has been defined, let me present the design model used for creation of TAUOLA C++ interface. Keeping in mind restrains explained in Section 2.2, the software would work as a bridge between TAUOLA FORTRAN as well as other software created by independent users and the current generation of event records. The design of the starting version for abstract algorithm was based on study of its FORTRAN predecessor.

The role of the interface will be to prepare information on the $\tau$ (four-momentum, spin state) in a format, which is understood by TAUOLA FORTRAN and as a post processing step to return (insert) $\tau$ decay products to the primary event record. Finally, role of such interfacing code will be to calculate dedicated weights from the production process information as well as from the decay, and unweight accordingly to standard MC procedures.

Accordingly, the model of this interface will be build upon three separate blocks: C++ interface to FORTRAN routines, abstract structure of the event record interface and implementations of different event records.

The original TAUOLA has been written for HEPEVT event record. The trick to connect algorithms based on this event record with the C++ one is to override the methods responsible for event record I/O with their C++ counterparts. Assuring that the C++ code will be executed instead of the old FORTRAN routines, no changes are required to FORTRAN code. This trick allows old algorithms to work as thought they use HEPEVT while the actual data from and to event record will be fed by the interface, collected from a C++ data structures. In case of TAUOLA, this solution becomes pretty simple as only `filhep_` method needs overwriting.

Starting from the FORTRAN code, we have defined the most basic needs for the interface and the decay algorithm. Using this approach we have build basic abstract structure and began planning the overall project. Following schema observed in HepMC and used previously in MC-TESTER, we have decided to divide algorithms regarding the whole event record and individual particles into separate classes – `TauolaEvent` and `TauolaParticle`. Additionally, knowing that the spin correlation algorithms will require information about pair of generated particles, we have created `TauolaParticlePair` to deal with all the information needed for single decay. The following two subsections encapsulate the most important information presented in the technical part of reference [1].

### 3.1 Software structure

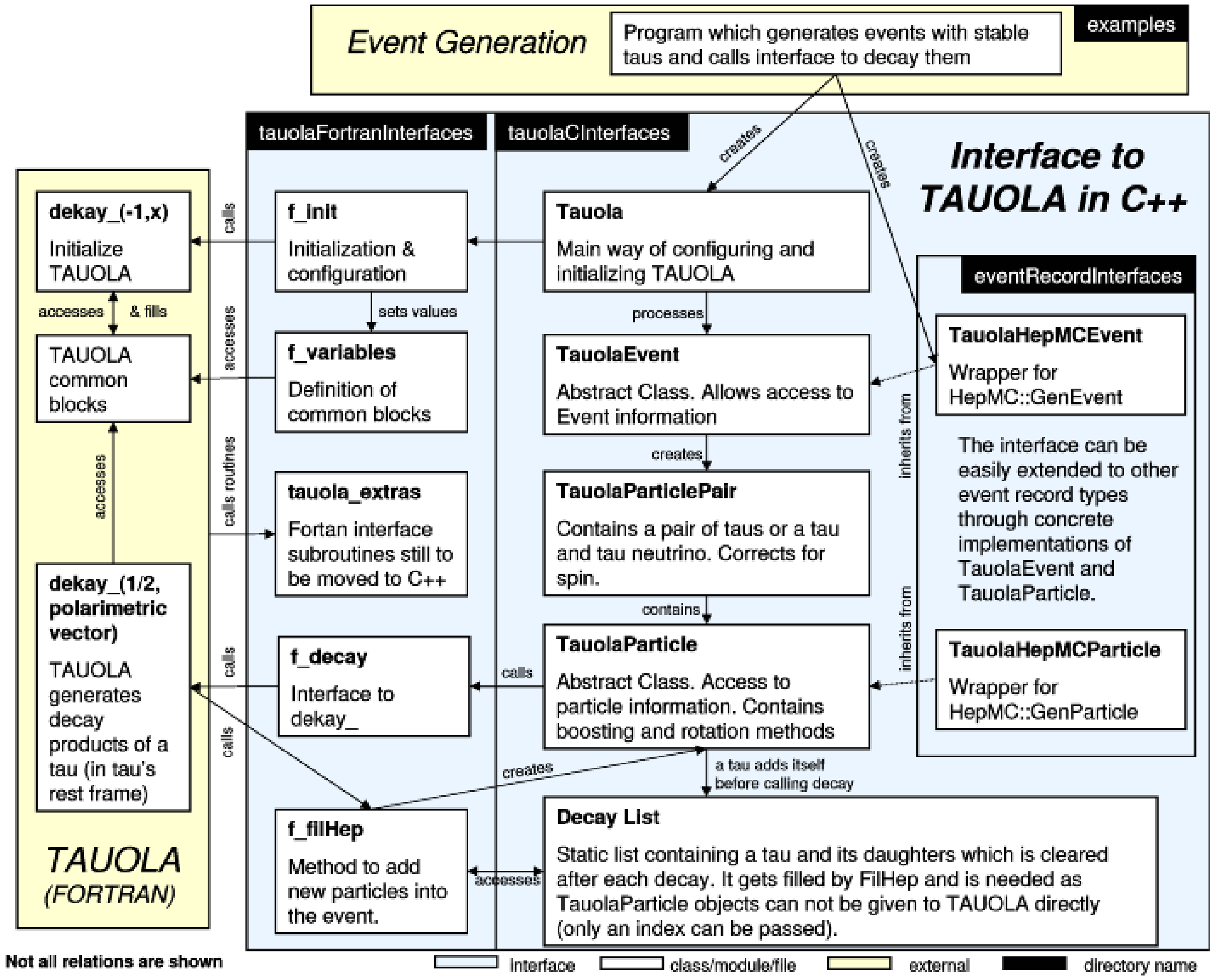

*Fig. 2 TAUOLA C++ Interface class relation diagram including TAUOLA FORTRAN and HepMC interface. Classes regarding the interface with electroweak library, random number generators and other methods are not shown.*

The choice of splitting the source code into three main modules, see *Fig. 2* (blue part), allows to separate the FORTRAN related code from the abstract C++ interface and the concrete implementation of the interface created for the appropriate event record.

- TAUOLA FORTRAN Interface
  This part of the code provides an interface to the FORTRAN library of TAUOLA. In particular, routines necessary for library initialization and wrapper for routine invoking the decay of a single τ. Parts of the interface code are still left in FORTRAN, but can be rather easily rewritten to C++. The most important method, `filhep_`, has been implemented in C++. Its FORTRAN predecessor writes single particles to the HEPEVT common block. At present, the method `filhep_` inserts the particle into the HepMC event record but remains to be called from the FORTRAN library.
- TAUOLA C++ Interface
  The abstract part of the interface to the event record. The class `TauolaEvent` contains information regarding the whole event structure, while `TauolaParticle` stores all information regarding a single particle. All particles used by the interface are located in the event in the form of a list of `TauolaParticle` objects. The last class located here, `TauolaParticlePair`, is the core of all polarization and decay algorithms. They are independent from the event record used by the interface as they operate on these two abstract classes presented above.

- Event Record Interface
  The event record implementation classes. All classes stored here represent the implementation of specific event record interfaces and are responsible for reading, traversing and writing to the event record structure. Only `TauolaEvent` and `TauolaParticle` classes must be implemented. The `HepMC` event record interface is implemented through `TauolaHepMC`Event and `TauolaHepMCParticle`.

By dividing the code into these three blocks, all new physics effects introduced independently from `FORTRAN` generation procedure can be easily implemented into abstract model of the generation process. Spin effects, electroweak corrections and effects of anomalous couplings can be introduced in this way.

## *3.2* Algorithm outline

The main algorithm used in the project consists of several steps starting from gathering information from event record through main τ generation to writing decay to event record.

Documentation of the `TAUOLA FORTRAN` Interface [16] describes some aspects of the spin correlation algorithm, which are also relevant to this interface.

- The `HepMC` event record is traversed and a list of all stable τ's in the event is created.
- From each found τ location, the tree is traversed backwards so that information about the production process can be extracted and used for the calculation of the spin density matrix.
- The siblings of the τ are identified through common parents. In cases such as $\tau \to \gamma\ \tau$, the parent(s) are defined as the particle(s) that produced the first τ; τ and $\nu_\tau$ siblings are paired to the τ.
- The density matrix is set-up using information about the τ-pair and their parent type (for Z/γ processes, grandparent information is also required).
- The pair is then decayed by executing the `FORTRAN` routine `dekay_` for each τ in the pair.
- A spin weight is calculated using the polarimetric vectors returned from `TAUOLA FORTRAN` and the density matrix previously set-up.
- If the decays are rejected, the pair is decayed anew and the process is repeated until the decays are accepted. In this way unweighting of spin effects is performed.
- Once accepted, the decay products are added into the event record with the procedure as follows:
  - The τ-pair are boosted and rotated into this hard process frame.
  - The `dekay_` routine of `TAUOLA FORTRAN` is executed with state = 11 or 12 (write). This initiates `TAUOLA FORTRAN` to return the daughter information via the `filhep_` routine.
  - The τ's status code is changed from stable particle to intermediate particle.

- A new particle object is created for each daughter and the appropriate tree structure is created and added into the event.
   - Each daughter is boosted using the τ's 4-momentum (as TAUOLA constructs decay for a τ at rest) to the hard process frame.
   - The τ's and their decay products are boosted back into the laboratory frame.
- As the final step, the position of vertices containing the τ's and their decay products is set according to the τ's momentum and lifetime.

The event is hence modified with insertion of τ decay products.

## *3.3* Decay tree structure

A typical structure of High energy physics process is represented by a tree of decaying particles. The decaying particle is considered mother of its decay products. Unstable daughters of the mother particle can decay further until stable particles are produced. Such interpretation corresponds to the tracks observed in the detector. In other parts of the tree it represent dominant, but not exclusive, Feynman diagram. The main algorithm of the interface regards the branch of the whole decay tree containing the decaying τ.

The data structures used in FORTRAN were focused on particles as the basic information source. Each particle stored information about itself and its relation with each particle. HepMC group particles into vertices representing single particle decay (see Section 2.4). Such organization makes it easier to identify decaying process of interest and overcomes limitations present in FORTRAN predecessors. However, with the new approach comes new set of exceptions that need to be dealt with, as the tree itself is prone to many assumptions and abbreviations that makes searching for mothers and grandmothers of intermediate particle a hard task.

First of all, the decay tree stored in event record may not have the tree-like structure at all. As indicated previously in Section 2.5, each Monte Carlo generator has its own way of storing information about different processes. Similar, data extracted from the detectors themselves might be corrupted as well. Several nodes might become disjoint due to lack of information and the decays themselves might have connections with completely separate parts of the tree. There are also cases when the background information regarding the process is stored in the event record as a separate, parallel tree. The tree becomes more of a graph and regular algorithms traversing the event record fail to overcome difficult exceptions. Structures observed in Monte Carlo generators can be divided into several categories:

Picture on the right presents the easiest type of event, where all information about hard process is explicitly given and there are no troubles finding mother and grandmothers of τ. It can be frequently observed in Monte Carlo generation process and is useful for test purposes. Unfortunately, such perfect event is rarely present in detector output.

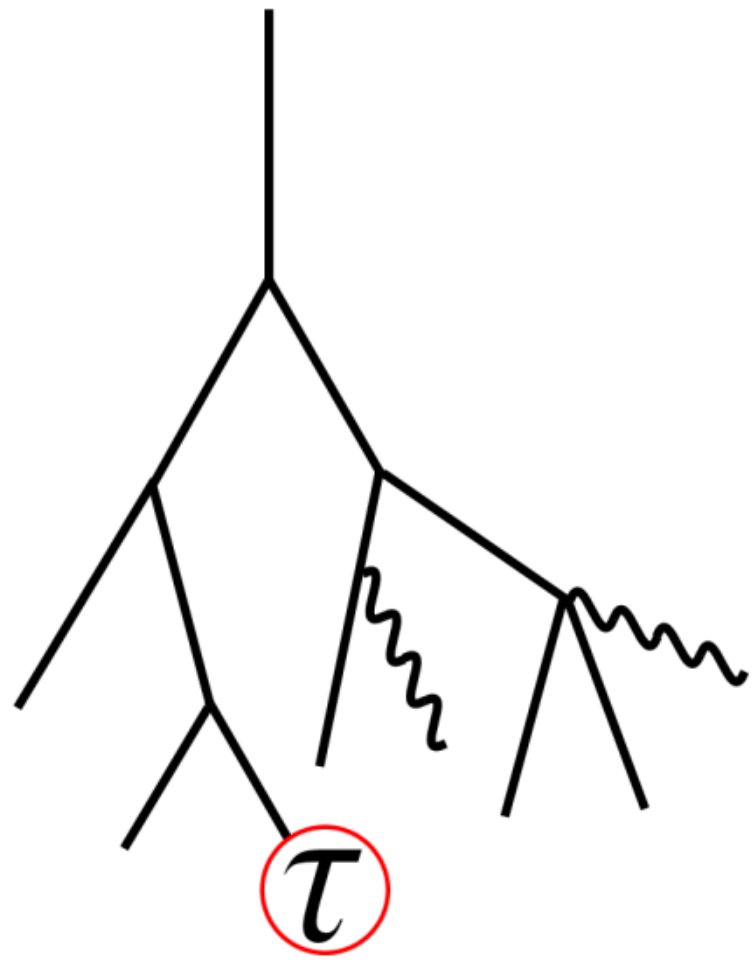

An example of $\tau \rightarrow \tau$ and $\tau \rightarrow \tau\gamma$ decay in `PYTHIA` Monte Carlo generator, in which the momentum is not conserved. However, this non-conservation is balanced, for example, between two branches outgoing from Z.

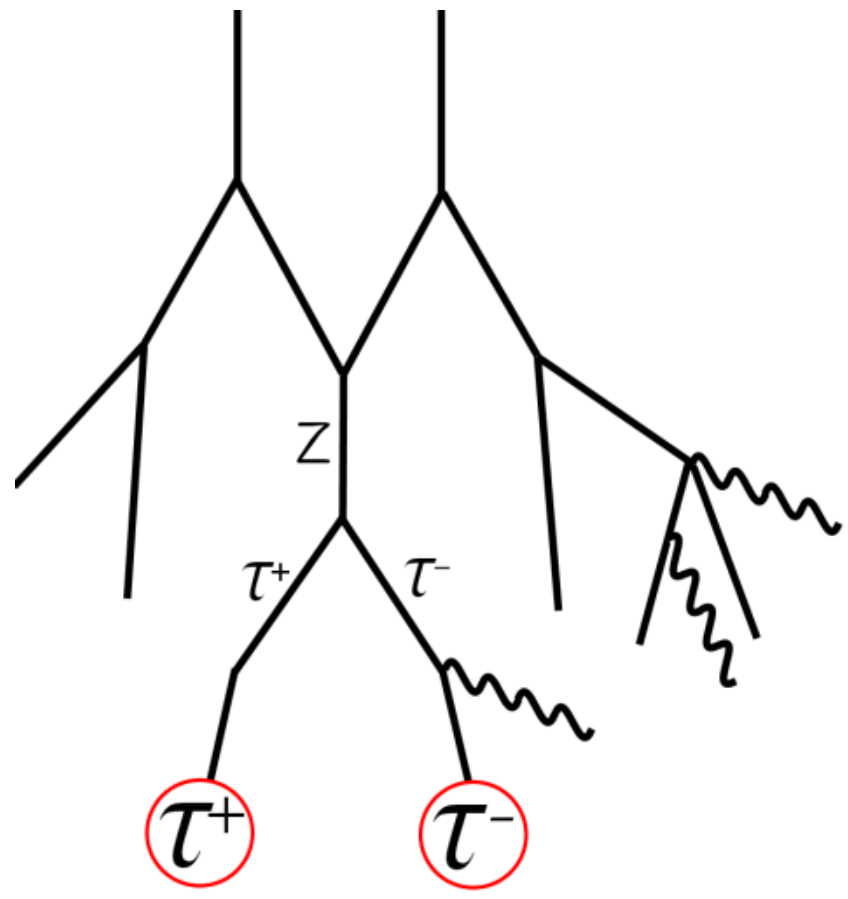

An example of a case when grandmothers are not explicitly given. In such cases, the interface is expected to reconstruct grandmothers of τ based on its mothers and siblings. Solution is not always deterministic.

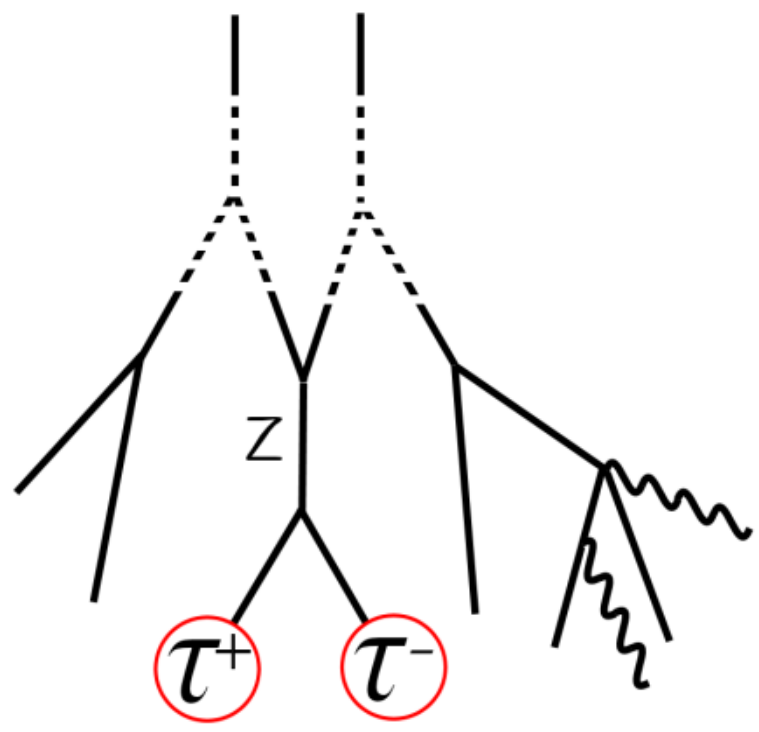

In general branching 2 → 1 → 2 is rare. Some cases of τ production may look like the picture on the right. Such cases imply many possible options and are source of many potential traps. The mother of τ needs to be constructed based on its siblings and grandmothers and the kinematics of hard process needs to be reconstructed as best as possible with the data given by event record.

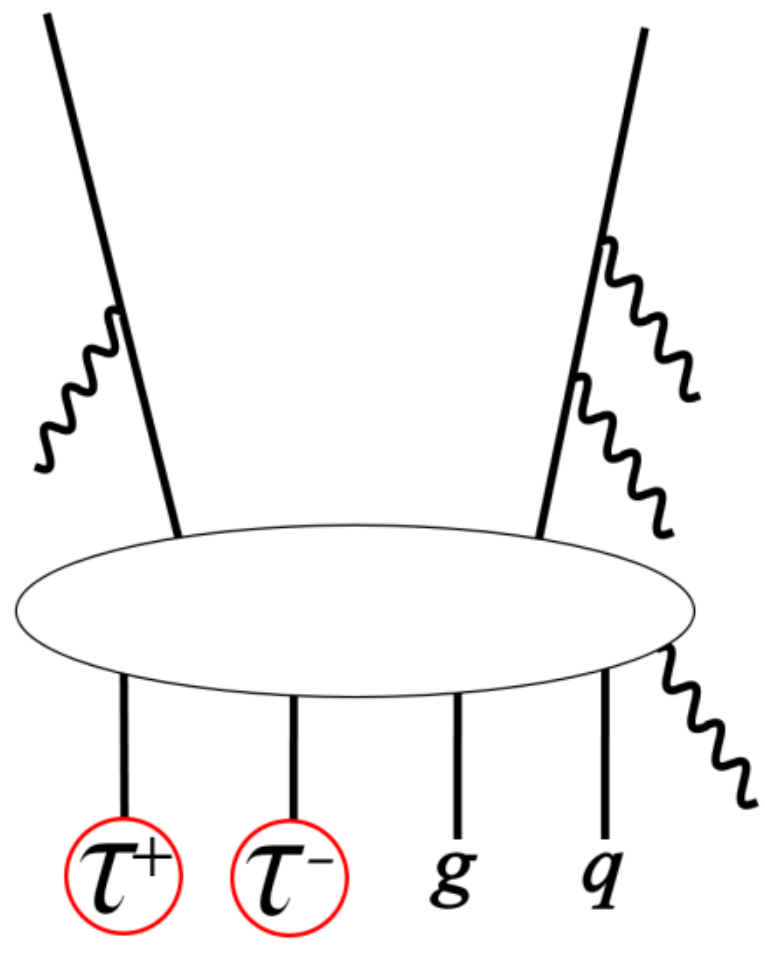

# 4. Implementation

Coordination and role separation within the project proved to be an essential part of the planning. Since the project is spanning over the vast amount of knowledge regarding High energy physics, my first step was to divide this knowledge into several separate block and decide what I had to learn, what might be useful to know, and with what I do not need to be concerned.

It came to my understanding that large amount of physics I do not have to understand to be able to code the necessary algorithms. This result in a fact that there are parts of the code that I perfectly understand and know how they work, but have no or very little knowledge about their meaning. It does not come natural to write code not knowing its actual meaning but I have realized that this is what is expected from a programmer working with large high energy physics projects as well as any other large industrial project. Having first tasks assigned and the roles in our project decided, we began implementing the abstract model following the outline described in previous chapter.

## *4.1* Methodology

Starting our project, we were not expecting how fast the standard within the collaboration will change. Having first part of the code ready, we already got informed that the new version of `HepMC` has been released introducing few significant changes [23] and forcing us to rewrite part of our code completely.

We were not too concerned with the changes of `HepMC` itself, as they could be dealt with thanks to separation of the abstract interface from the concrete implementation. The real problem was that these changes resulted in different representations of the physics process itself and as the consequence they require change of the abstract level of the algorithm. Such changes imply that software has to be rearranged. Since these changes regard the basis of the analysis – the event record content – short but non trivial modifications of the code may be necessary.

Such changes are host-generator dependent. If more than one interpretation is necessary (as in case of PYTHIA 8.1 and 6.4, for example), it may result in multiple implementations. The interface must take into account all of these options as well as the fact that some experiments might not incorporate new versions of physics Monte Carlo generators as quickly as the others might. Compatibility with the older versions must exist as well.

We have quickly realized that regular approach to software design fails, as it does not account for such fundamental changes lying completely outside of our project. It became obvious that since the software must be compatible with collaboration standards, it means that it must be prepared to the frequency of changes introduced within collaboration. Therefore, we have changed our methodology to iterative one, based on agile software development [24], focusing on response to changes rather than following a plan.

The development cycle in most popular methodology for well-defined project – the waterfall approach – starts with definition of requirements, followed by creating the design, implementing the abstract model, testing and maintaining the final software.

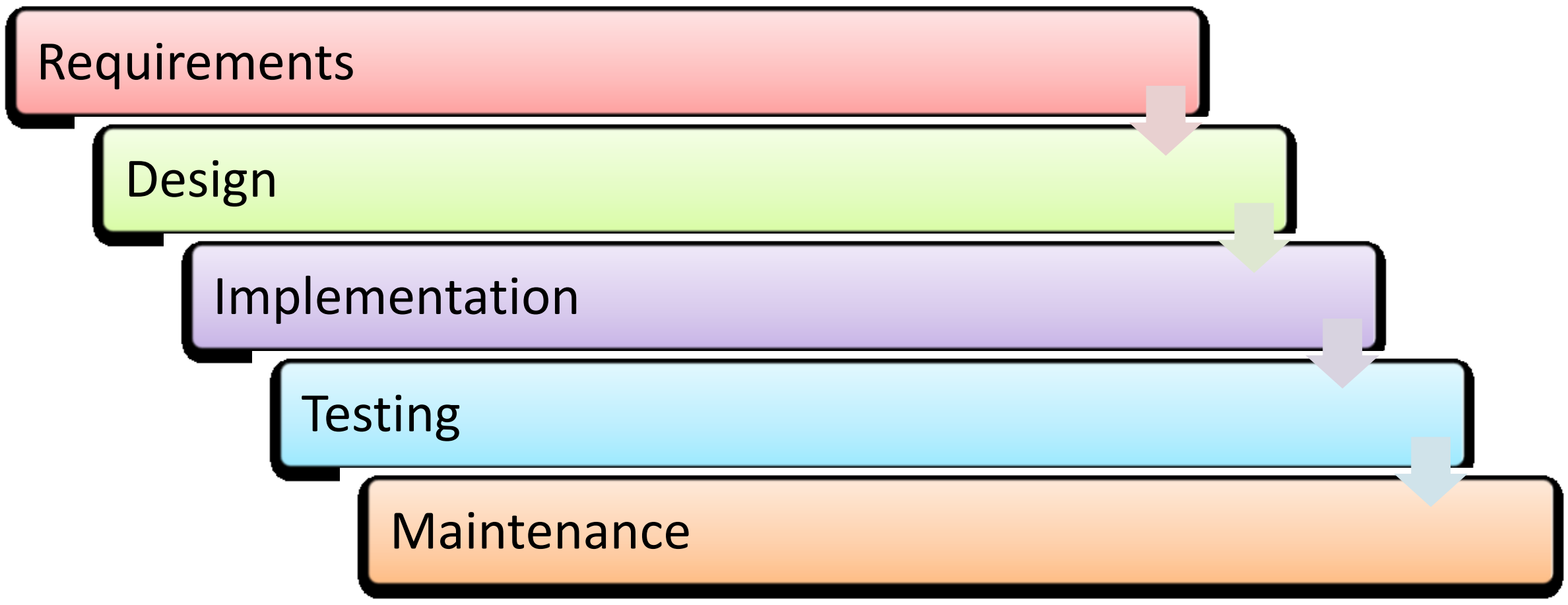

*Fig. 3 Five steps of waterfall approach*

If the specification was detailed enough, and assuming that the communication between the team and the client was as best as possible, creating a design that suits customer the most, then the finished product is expected to have full functionality declared by the customer, will be made within the budget and will be delivered on time. However, the main issue of high energy physics software is the fact, that the main prerequisite for waterfall approach to be successful cannot be fulfilled. It is impossible to plan full functionality of the finished project. The basics of the project are already defined and can be easily covered in more details, but because the software itself is not made for any specific client, organization or specific experiment, with multitude of different uses there is no way to tell what functionality end users might request.

In case of rapidly changing projects, where the functionality is created on the fly, an iterative approach has been designed and introduced in form of agile software development manifesto [25]. This approach presents a completely different view on software development cycle. It starts with the basic planning followed by a repetition of iterations in form of three steps – iteration planning, execution and evaluation.

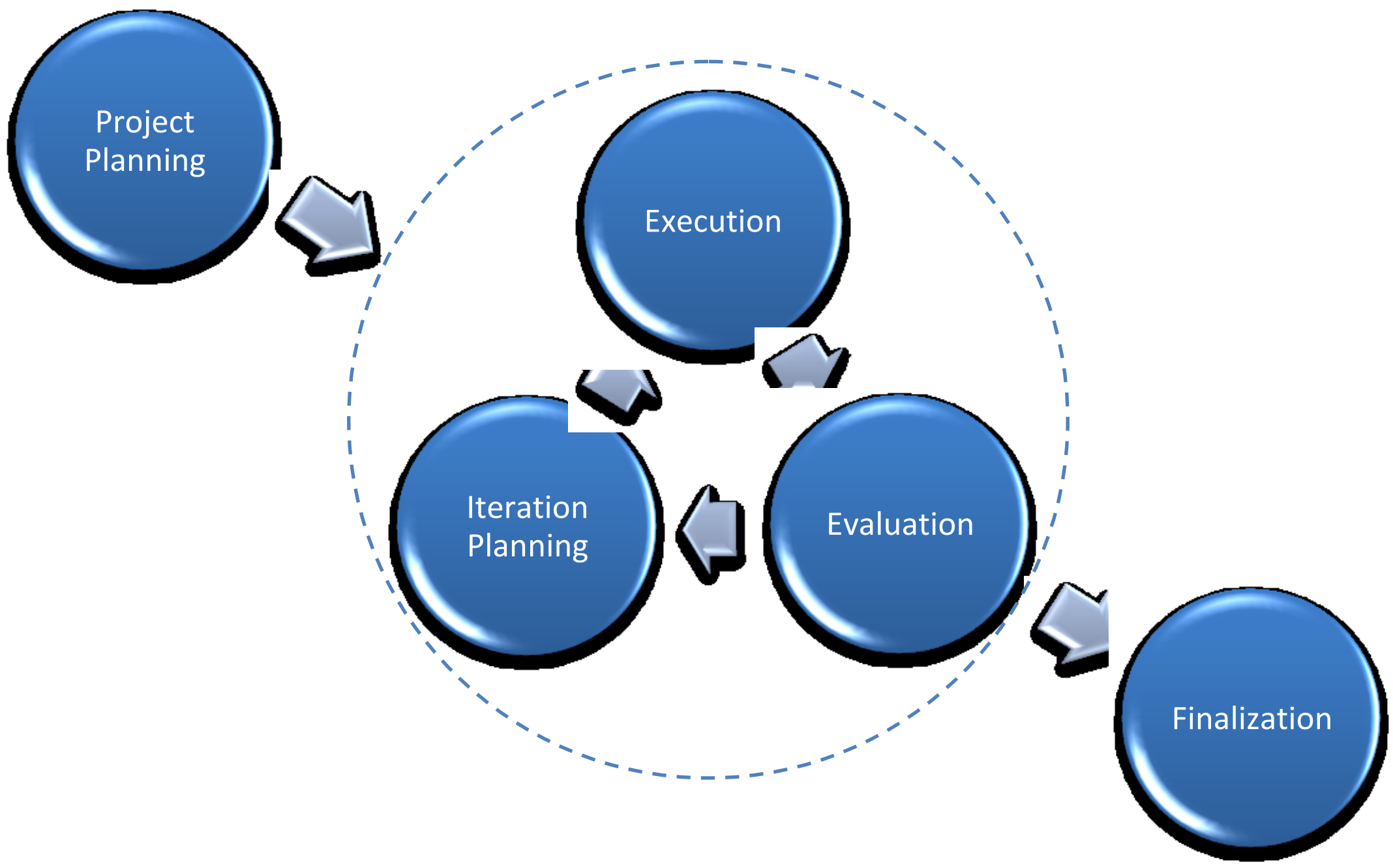

*Fig. 4 Iterative approach of agile methodology*

Within each cycle, a new functionality is implemented and the list of the next tasks created to be assigned in the upcoming iteration. This way each iteration results in the new, extended version of the previously written code. The cycle ends with the consolidation of the finished project. The differences are obvious – there is no need to define full functionality at the start of the project, but instead the customer is required to verify functionality created in each cycle and to provide tasks for the next one. One of the disadvantages of such loosely defined design is that the outcome of the whole development process cannot be fully predicted. It implies that the project can end with more functionality implemented than originally expected at the cost of extended development time or budget. It can also mean that the project might be finished on time and within the budget, but with less functionality than expected. Additionally, since the customer does not fully define the project at the very beginning, his attention is required throughout the whole development process.

The agile methodology was created specifically to deal with the problems that regular methodology cannot solve. However, it has many requirements that high energy physics projects cannot comply. For starters, it requires good management and well-coordinated team to be fully successful. Writing software for physics experiments requires cooperation with specialists around the world and implies that some of the team members can join in during project development while others may leave the project due to lack of time or other issues that require their attention. The other problem is, as mentioned before, constant interaction with the users, which in case of high energy physics projects is hard to establish. The agile methodologies value interaction with customer in terms of planning the new functionality. However, as writing project for High energy physics experiment is different from the commercial software, it is hard to tell which experiments except ATLAS and CMS might actually be interested in the software as well as what potential user might expect. The requirements for new iteration may appear unexpectedly as some of the users find functionality that they would like to be implemented, while for the most of the time user feedback may be simply nonexistent.

In search for methodology that would suit our project the most, we started from a popular in our country methodology called XPrince [26]. Based on a software management methodology, as well as an agile approach, XPrince is a hybrid methodology defining stable steps of software evolution and means to adopt the software to constantly changing requirements. As an alternative approach, we have also tried commercial software offered by Parasoft. `Parasoft Concerto` [27] is agile software development management application designed to help planning, defining requirements and managing tasks allocated to different team members. It supports both waterfall and agile approach including hybrid methodologies as well as mixture of these types within groups working on the same project. We were not able to test this solution during development of `TAUOLA` C++ Interface; however, we considered its usefulness in projects with larger team requiring much more planning and cooperation.

Since the most basic requirements of our project could be well defined, we were able to plan and start implementation of the core functionality of the project. As the first versions became available and first tests were performed, we have started looking for testing environments. It was quite shocking how different experiments focus on their own, unique simulation software rather than following any common standard. It became obvious that our software will require specific arrangement for each experiment, forcing us to create specific fixes for each platform.

The model of our software started to change along with the code and new tasks were assigned based upon the previous iteration. Using doxygen, we have set up a website providing documentation, early software releases and necessary installation information for future testers. We have also set up an automatically generated release archive updated daily, which created a pace that forced us to quickly incorporate new modifications and always create a working version. Each day new version was available to the users.

Thanks to user feedback, with new releases came new functionality requests and changes to existing ones which sometimes required complete rewrite of some parts of the code. Before introducing new functionality to the software, we had to be sure that changes will not damage existing code. An extensive testing procedure had to be implemented so additional test modules were created including memory leak tracking function and internal algorithms testing plots as well as extended set of benchmark tests. We have also created several `MC-TESTER`-specific macros for advanced tests[3]. Having basic installation scripts prepared, the first stable version has been released.

## 4.2 Spin correlations

This section summarizes the implementation of the spin correlations algorithm used in the interface. The physics aspects of this algorithm are described in more detail in reference [1].

The basic algorithms for spin correlations already existed in `FORTRAN` code. However, we have decided to write it anew as from physics point of view, the code is easy to understand, while the old algorithm relied on the older event record; therefore, it was worth rewriting it to C++ as a part of the interface.

If more than one τ lepton is present in a final state, then not only is the individual spin state for each τ necessary for proper generation, but the complete correlation matrix of all τ leptons must be taken into account as well. In the case of τ-pair production, the standard algorithm explained in [13; 15] can be used without much modification. For the single τ produced in a $\tau - \nu_\tau$ pair, it is convenient to use the same algorithm as well, even though it is not necessary from the physics point of view.

As described in papers mentioned above, spin correlations and spin polarization effects can be simulated by accepting or rejecting a pair of generated τ decays based on a weighting factor `wt`.

$$wt = \frac{1}{4} \sum_{i,j=0}^{4} h_i^1 \; h_j^2 \; R_{ij}$$

where $h^1$ and $h^2$ are the polarimetric vectors for the $\tau^+$ and $\tau^-$ respectively and $R_{ij}$ is the density matrix associated with the τ production vertex. The matrix $R_{ij}$ depends on the mechanism and particular kinematical configuration of τ pair production. The $h_i^1$, $h_j^2$ depend on the respective decays of $\tau^+$ and $\tau^-$. The solution can be used for $\tau - \nu_\tau$ production as well. In this case, $\nu_\tau$ decay is not performed and its polarimetric vector is set to $h = (2,0,0,0)$.

[3] Results of these tests are presented in Section 4.1.

A pair of τ decays should be accepted if the weight is greater than a randomly generated number between 0 and 1. If this condition fails, the τ pair decays should either be rejected and regenerated, or rotated[4] and the weight recalculated. The production process does not need to be reprocessed.

In the formulas used to create matrix $R_{ij}$ the hard process kinematical variables $s$ and $\theta$ have to be known for each event. Those variables are also used by a module for calculating electroweak corrections (see Section 4.3).

Variable $s$ is defined as square of invariant mass of τ mother. To find the angle $\theta$ we need to identify the four momenta of $\tau^+$ and $\tau^-$ pair. Two scattering angles $\theta_1$ and $\theta_2$ can be reconstructed. The angle $\theta_1$ is between $\tau^+$ and the first incoming state, $\theta_2$ is between $\tau^-$ and the second one[5]. Both angles are calculated in the rest frame of the τ pair. The average angle $\theta^\bullet$ accordingly to the description given in [28] is taken:

$$cos\theta^\bullet = \frac{sin\theta_1\, cos\theta_2 + sin\theta_2\, cos\theta_1}{sin\theta_1 + sin\theta_2}$$

The density matrices $R_{ij}$ for the most standard processes of τ-pair production are documented in details in [1].

The algorithm for the spin correlations, in contrast to the other parts of the code, forces to look outside the single node in the decay tree. The gathering of information regarding parents and grandparents of the decayed particle requires traversing the event record backward to search for appropriate particles. However, this poses many problems as in most cases, the decay tree is not as perfect as the theory describes.

Due to the number of options of possible decay tree structures (see Section 3.3) a unique algorithm had to be written. The hard τ's are used to calculate spin correlation and the stable τ for kinematics, which complicates the procedure even further. The algorithm was tested and was modified using hundreds of different event record examples. As it became obvious that any new modules introduced must be easily replaceable in case of fundamental changes, further work on advanced spin effects was gathered into separate class: `TauolaParticlePair` (see Section 3.1). The algorithm is independent from the rest of the project and if data provided from event record is insufficient, no spin correlation will be calculated. Thanks to this approach, the code becomes more robust as it takes into account any type of event record, including damaged ones with incomplete or invalid tree structure.

## 4.3 Electroweak corrections

In some cases, notably in the case of $\tau^+ \tau^-$ produced from the annihilation of a pair of quarks, the standard density matrices may not be sufficient for some applications. A more exact solution is also available. Instead of a native $R_{ij}$ density matrix, an externally calculated one can be used.

---

[4] Rotation instead of rejection increases efficiency by a factor of 4. This however only affects the generation of τ lepton decays and represents a small fraction of the total time of constructing the event.,

[5] Assuming the first incoming state to be antiparticle.

The solution is based on SANC library [19; 29] for calculation of electroweak corrections[6]. With its help, the density matrix $R_{ij}$ for $q\bar{q} \to \tau^+ \tau^-$ process can be calculated as a function of the incoming state flavour and Born level variables (Mandelstam $s$ and scattering angle $\theta$). Additional two weights are also provided, which include the matrix elements squared and averaged over the spin. For additional weights, genuine weak corrections are respectively switched on and off[7].

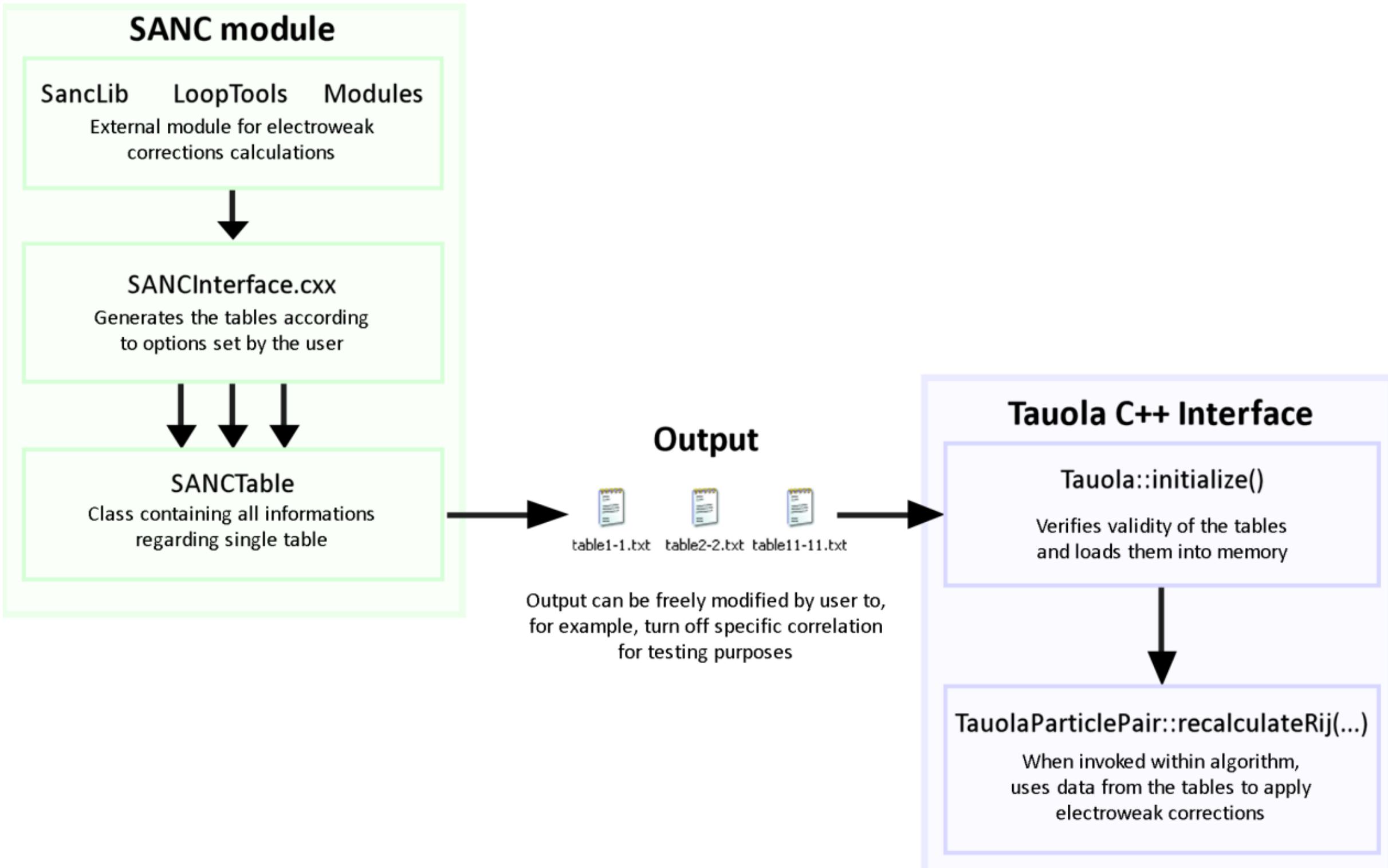

*Fig. 5 Organization of electroweak corrections module.*

Keeping modular approach in mind, and to speed up execution of the program, electroweak corrections has been implemented as a completely standalone unit. A separate program generates output in form of several tables that the interface can read. The tables calculated by this unit store the $R_{ij}$ values in a lattice of $(s, cos\theta)$ points.

Later, in the actual execution of our interface, these pretabulated values of $R_{ij}$ are interpolated to the actual phase space point. For this purpose, the standard bilinear interpolation algorithm is used. Additionally, in order to avoid numerical errors, for $cos\theta$ values near -1 and 1 we are using the linear extrapolation algorithm.

The advantage of this solution is that results of SANC library calculation can be modified by the user before it is loaded into our interface without intervention into the code of the interface itself. Pretabulation is prepared for three domains of s: around the Z peak, close to the

[6] It may serve as an example of how other calculations featuring heavy Z' boson, for example, may be used in our interface.

[7] This may be helpful for the evaluation of genuine weak corrections for states of large s, significantly above the Z peak, where they become sizable. See eg. refs [33],[34].

WW pair production threshold and over a broad energy range, but the actual choice of pretabulation zones can be easily modified by user. The algorithm outline of applying electroweak corrections is presented below:

- Get information about the production process
- If a table for such process does not exist, return the default value[8]
- Calculate $s$ and $cos\theta$ as described in Section 4.2
- If s is below designated threshold, return the default value
- Check which range of pretabulation is the most suited for interpolation
- Find the four points on the lattice closest to $(s, cos\theta)$
- Interpolate the value of electroweak correction

The program generating the tables can be easily adopted to account for new physics processes or to create a specific tests. For complicated cases or for advanced testing purposes, user might prefer to modify tables manually using external software or a script. Additionally, as in case of spin correlation, in absence of pre-generated tables, or in case these tables has been modified by user in an inappropriate way, the interface will ignore electroweak corrections returning the default value.

## 4.4 FORTRAN TAUOLA interface

As described in Section 3.1, the core algorithm for generating the τ decays has been integrated into the interface as an interchangeable module with several routines interfaced to C++ code:

- `filhep_` - as indicated in Section 3, this routine for event record I/O has been overridden by a new version written in C++. It has been adopted to use HepMC as the means of communication between FORTRAN and C++ part of the code, however in future it will be adapted to use all supported C++ event records.
- `dekay_` - the most important routine of TAUOLA FORTRAN. It is called to generate τ and return its polarimetric vector used in algorithm described in Section 3.2 as well as to generate daughters of τ and save all newly generated particles to event record.
- `taupi0_`, `tauk0s_`, `taueta_` - if appropriate option is used (see Section 4.5.5), these routines are invoked to generate decays of η, $K_S^0$ and $\pi^0$ respectively. This allows short living scalar particles produced in τ decays to be decayed by the interface rather than by invoking a host generator after τ, decays are produced.

The rest of the interfaced code consists of the list of common blocks used for setting up options for TAUOLA FORTRAN (for more details see reference [1]). Several routines (such as `plzap0_`) have been left in FORTRAN, but have been statically included into the FORTRAN

[8] As the default value, the analytic form taken from ref [41] is used. It features all spin and mass effects, but electroweak corrections, and even Z exchange, are not taken into account. This is reasonable for $s<36GeV^2$

interface. They represent parts of old FORTRAN code for calculation of spin correlations and interface to module for calculation of electroweak corrections. They are prepared to be re-written to C++ in near future. As their use poses no difficulty, they are left as such. In principle, they offer an example of numerically better solution than pretabulation of electroweak corrections method used by us. They operate at the level of form factors used in spin amplitudes calculations instead of the level of density matrix and cross section. Such functions are by far smoother.

## 4.5 User configuration

For purpose of configuring the decays generated by the interface, a new module has been created in form of a static class `Tauola`. This class stores configuration options for each step of the simulation. Configuration of both TAUOLA FORTRAN and the interface itself can be performed using options divided into three categories – initialization, generation and output.

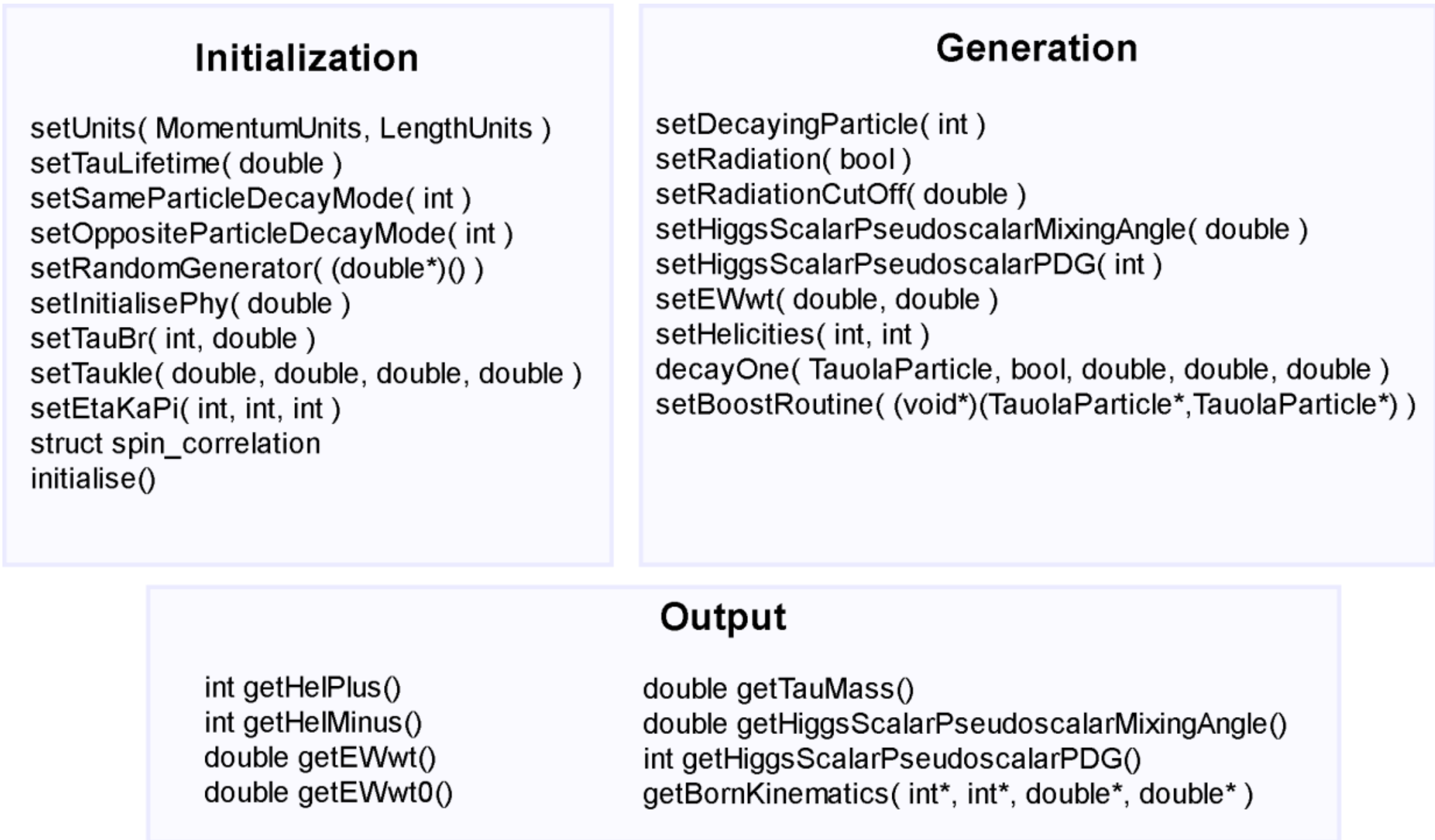

*Fig. 6 Configuration options of TAUOLA C++ Interface.*
*This list includes options described in Section 6.4.*

This section contains description of several options as an example of functionality implemented during project development. For details regarding rest of the options, see reference [1].

### 4.5.1 Decaying Particle

The following method is prepared to impose the sign for the "first τ", that is to reverse signs of `SameParticle` and `OppositeParticle` used in decay mode selection (see section below):

- `Tauola::setDecayingParticle(int pdg_id)`
  Set the PDG id of the particle which `TAUOLA` should decay as “first τ”. Both particles with pdg_id and -1*pdg_id will be decayed. Default is 15, one may want to use -15 for special applications.

Example:

`Tauola::setDecayingParticle(-15);`

*Set SameParticle to be $\tau^+$*

### 4.5.2 Decay mode selection

By default, all τ decay modes will be generated according to predefined branching fractions. Methods to modify the default values are provided:

- `Tauola::setSameParticleDecayMode(int mode)`
  Set the decay mode of the τ with the same PGD code as set by setDecayingParticle()
- `Tauola::setOppositeParticleDecayMode(int mode)`
  Set decay mode of the τ with the opposite PGD code as set in setDecayingParticle()

Example:

`Tauola::setSameParticleDecayMode(Tauola::PionMode);`

`Tauola::setOppositeParticleDecayMode(4);`

*Forces only the modes $\tau^- \to \pi^- \nu_\tau$ and $\tau^+ \to \rho^+ \nu_\tau$ $(\rho^+ \to \pi^+ \pi^0)$ to be generated*

- `Tauola::setTauBr(int mode, double br)`
  Change the τ branching ratio for channel `mode` from default to `br`.

Example:

`Tauola::setTauBr(3,2.5);`

*Sets rate for channel $\tau^\pm \to \pi^\pm \nu_\tau$ to 2.5. All channel rates may not sum to unity, normalization will be performed anyway.*

The int mode enumerators which are arguments of the above functions have the following meaning:

- 0 – `Tauola::All` - All modes switched on
- 1 – `Tauola::ElectronMode` – $\tau^\pm \to e^\pm \nu_\tau \nu_e$
- 2 – `Tauola::MuonMode` – $\tau^\pm \to \mu^\pm \nu_\tau \nu_\mu$
- 3 – `Tauola::PionMode` – $\tau^\pm \to \pi^\pm \nu$
- 4 – `Tauola::RhoMode` – $\tau^\pm \to \rho^\pm \nu$
- 5 – `Tauola::A1Mode` – $\tau^\pm \to A_1^\pm \nu$

- 6 – `Tauola::KMode` – $\tau^{\pm} \to K^{\pm}\nu$
- 7 – `Tauola::KStarMode` – $\tau^{\pm} \to K^{*\pm}\nu$
- 8 – $\tau^{\pm} \to 2\,\pi^{\pm}\pi^{\mp}\pi^{0}\nu$
- 9 – $\tau^{\pm} \to 3\pi^{0}\pi^{\pm}\nu$
- 10 – $\tau^{\pm} \to 2\,\pi^{\pm}\pi^{\mp}2\pi^{0}\nu$
- 11 – $\tau^{\pm} \to 3\pi^{\pm}2\pi^{\mp}\nu$
- 12 – $\tau^{\pm} \to 3\pi^{\pm}2\pi^{\mp}\pi^{0}\nu$
- 13 – $\tau^{\pm} \to 2\,\pi^{\pm}\pi^{\mp}3\pi^{0}\nu$
- 14 – $\tau^{\pm} \to K^{\pm}K^{\mp}\pi^{\pm}\nu$
- 15 – $\tau^{\pm} \to K^{0}\overline{K^{0}}\pi^{\pm}\nu$
- 16 – $\tau^{\pm} \to K^{\pm}K^{0}\pi^{0}\nu$
- 17 – $\tau^{\pm} \to 2\pi^{0}K^{\pm}\nu$
- 18 – $\tau^{\pm} \to \pi^{\pm}\pi^{\mp}K^{\pm}\,\nu$
- 19 – $\tau^{\pm} \to \pi^{\pm}\pi^{\mp}\overline{K^{0}}\,\nu$
- 20 – $\tau^{\pm} \to \eta\pi^{\pm}\pi^{0}\nu$
- 21 – $\tau^{\pm} \to \pi^{\pm}\pi^{0}\gamma\nu$
- 22 – $\tau^{\pm} \to K^{\pm}K^{0}\nu$

### 4.5.3 Spin correlations options

By default, all spin correlations are turned on. However one may be interested to partially or completely switch off their effects for the sake of numerical experiments which validate whether a measurement will be sensitive to certain spin correlation components. This technique may be useful to evaluate the significance of spin correlations for signal/background separation as well. For this purpose the following methods are provided:

- `Tauola::spin_correlation.setAll(bool flag)`
  Turns all spin correlation computations on or off depending on the flag, which can be either true or false. Note: this should be called after `Tauola::initialise()`.
- `Tauola::spin_correlation.HIGGS=flag`
  Turns particular spin correlation computation on or off for a given τ parent depending on the flag which can be either true or false. Implementation of this switch is provided for: `GAMMA, Z0, HIGGS, HIGGS_H, HIGGS_A, HIGGS_PLUS, HIGGS_MINUS, W_PLUS, W_MINUS`. The keywords denotes the τ parent.

Example:

```
Tauola::spin_correlation.setAll(false);
Tauola::spin_correlation.HIGGS=true;
```

*Turns all spin correlations off, except HIGGS.*

### 4.5.4 Radiative correction

The user may want to configure parameters used in the generation of QED corrections in the leptonic decay channels of τ's. For that purpose the following methods are provided:

- `Tauola::setRadiation(bool switch)`
  Radiative corrections for leptonic τ decays may be switched on (default) or off by setting the switch to true or false.
- `Tauola::setRadiationCutOff(double cut_off)`
  Set the cut-off for radiative corrections of τ decays. The default of 0.01 means that only photon of energy (in its rest frame) up to 0.01 of half of the decaying particle mass will be explicitly generated.

Example:

```
Tauola::setRadiation(false);
```

*Switch radiative corrections off in τ decays*

### 4.5.5 Decay of final state scalars

In some cases a user may want TAUOLA to decay short living scalar particles produced in $\tau^{\pm}$ decays, rather than invoking a host generator for the post processing step. For that purpose a special algorithm is prepared, even though high precision is then not assured. This might not be a problem if the algorithm is used for τ decays only where events with such decays are rather rare:

- `Tauola::setEtaK0sPi(int a, int b, int c)`
  Three parameters `a`, `b` and `c` switch on or off the decay of η, $K_S^0$ and $\pi^0$ respectively. A value of 1 is on and 0 is off.

Example:

```
Tauola::setEtaK0sPi(1,0,1);
```

*In event branch starting from τ, η and $\pi^0$ decay, but $K_S^0$ remains undecayed.*

### 4.5.6 Helicity states and electroweak correcting weight

Independent of the generation process, the information on helicities of $\tau^+$ and $\tau^-$ can be returned[9] with the help of accessors:

```
int Tauola::getHelPlus()
int Tauola::getHelMinus()
```

Note that these helicities are not used in the interface and carry approximate information only. The electroweak weight can be returned with the help of accessors:

---

[9] The actual sign may depend on the process and boosting routine. Approximations in attributing helicities are introduced.

```
double Tauola::getEWwt()
double Tauola::getEWwt0()
```

These methods provide information once processing of a given event is completed.

## 5. Numerical tests and physics results

Creating a solid testing environment is an integral part of the development process. As new functionality is introduced, new tests needs to be created and the code needs to be validated again before any progress can be made. As soon as the validation of each version of the code ends, new tests can be created to verify less visible effects or to enhance testing procedure.

Starting from the tests created for number of FORTRAN projects (including TAUOLA FORTRAN), our testing evolved throughout the whole development of the interface. It was systematically extended to account for all implemented functionality. For this purpose, all FORTRAN benchmark tests had to be rewritten to C++ basing only the concept of the previous tests. It was, nonetheless, a significant help in creating the fundamental part of the tests.

Furthermore, the increase of the level of precision used changes conditions for tests. On one side, increased precision help to extend program stability and application domain as numerically small effects in one application confirm that there will be no bigger problems for other cases. On the other hand, creation of such tests becomes more complicated as small effects are usually more difficult to pin down.

Due to complexity of the project, there are many cases in which regular user might not be able to deduce how the final distributions of different aspects of τ decays might look like. Therefore, it is crucial that the user is able to reproduce these distributions and to study their individual aspects. Following section contains the most significant results published in [1]. The results presented here serve as examples of how both validation of existing processes and observation of new effects can be achieved.

The results of these tests were compared to the results obtained with the FORTRAN Interface (which has been well validated by comparison with analytical and numerical calculations for τ pair production[10]. For a review of physics oriented tests of τ decays themselves, and projects for future improvements based on low energy $e^+e^-$ data, see ref. [3]).

In addition to this, we created custom MC-TESTER macros for plotting other spin sensitive quantities and compared these to published results. Numerical results are presented later in the section see *Fig. 9a*, *Fig. 9b* and *Fig. 10a*, *Fig. 10b*.

### *5.1* Basic tests

The most basic test, which should be performed, is verification that the interface is installed correctly, that all τ leptons are indeed decayed by the program and that energy momentum conservation is preserved. TAUOLA has its own database of parameters and as a consequence the τ lepton mass may differ between the program performing a τ's production and

[10] This represents tests of interface. In all cases τ decays are generated with the help of TAUOLA FORTRAN

TAUOLA performing its decay. This leads to the sum of τ decay product momenta not exactly matching the τ's momentum. Although this effect may seem negligible, it may break numerical stability of programs like PHOTOS if they are applied later.

Once correct execution of the basic program steps have been confirmed, i.e. τ leptons are decayed, energy momentum is conserved and there are no double decay occurrences in the event tree, step one of the program installation tests is completed. In principle, these tests have to be performed for any new hard process and after any new installation. This is to ensure that information is passed from the event record to the interface correctly and that physics information is filled into event record in expected manner. Misinterpretation of the event record content may result in faulty generation by TAUOLA. For example, spin correlations may be missing or badly represented, or some τ leptons may remain undecayed.

## 5.2 $Z/\gamma \to \tau^+ \tau^-$

The longitudinal spin effects for Z decay into τ's was tested by restricting the τ decay mode to $\tau^\pm \to \pi^\pm \nu_\tau$ and examining the invariant mass of the $\pi^+\pi^-$ pair, $M_{\pi^+\pi^-}$ (see *Fig. 7a*) and the π energy distribution in the rest frame of the Z (see *Fig. 7b*). The effect of Z polarization on these distributions was studied in [30] and we obtained consistent results with the new C++ implementation of the TAUOLA Interface.

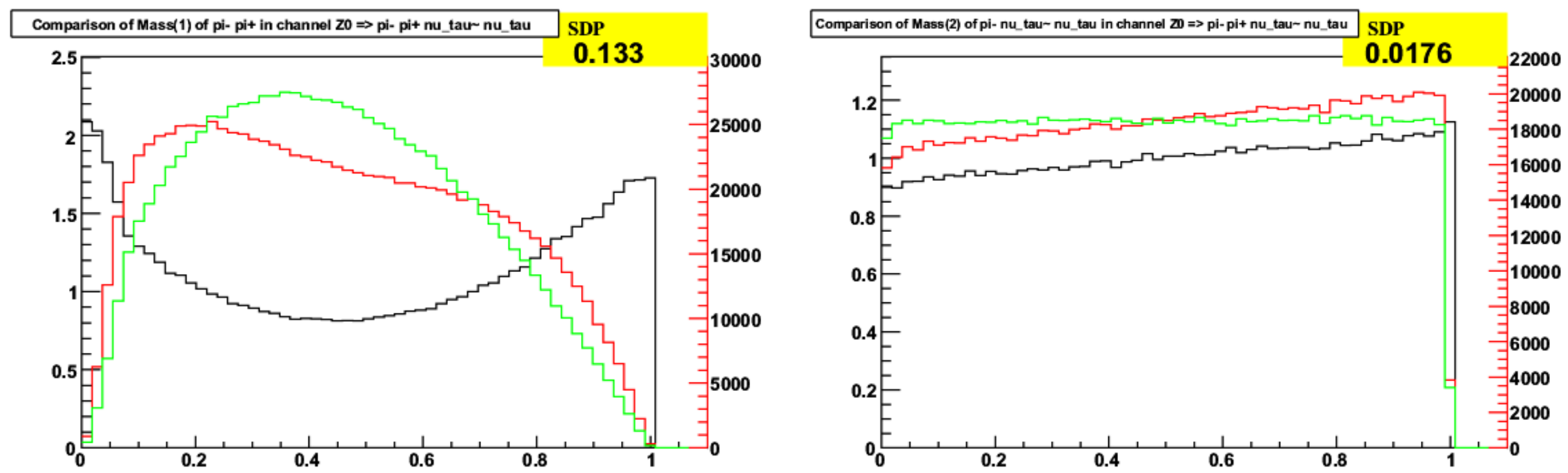

a) $M_{\pi^+\pi^-}$ b) $1 - 2\frac{E_{\pi^+}}{M_Z}$

*Fig. 7 Longitudinal spin observables for the Z boson (*$\boldsymbol{e^+e^-} \to \boldsymbol{Z}$ *at 500 GeV). Distributions are shown for spin effects switched on (red), spin effects switched off (green), and their ratio (black)*

## 5.3 $H^0/A^0 \to \tau^+\tau^-$

As was done for Z decay in previous section, longitudinal spin effects for Higgs decay into τ's was tested using $M_{\pi^+\pi^-}$ (*Fig. 8a*) and the π energy distribution in the rest frame of the $H^0$ (see *Fig. 8b*), which was flat as expected.

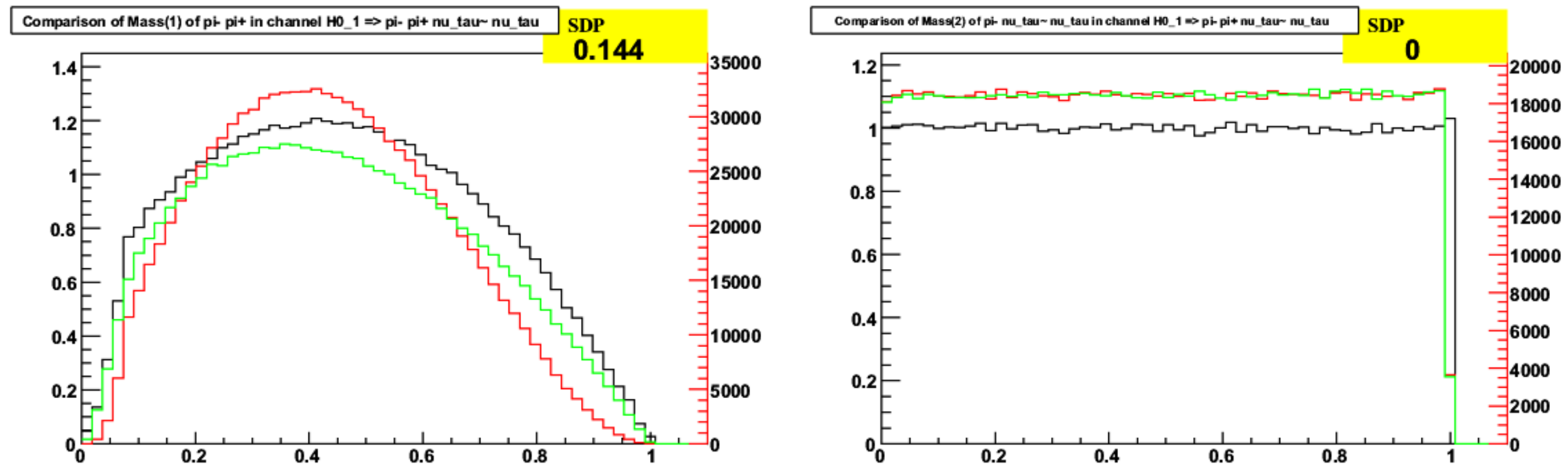

a) $M_{\pi^+\pi^-}$ b) $1-2\frac{E_{\pi^+}}{M_Z}$

*Fig. 8 Longitudinal spin observables for the H boson for* $\boldsymbol{\tau^\pm \to \pi^\pm \nu_\tau}$*. Distributions are shown for spin effects switched on (red), spin effects switched off (green), and their ratio (black)*

Let us now turn to transverse spin correlations. In *Fig. 9a* the benchmark histogram as produced by our FORTRAN Interface and given in Fig. 3 of reference [31] is reproduced[11]. It features acollinearity of the $\boldsymbol{\pi^+\pi^-}$ pair in the Higgs boson rest frame, both τ's decay to $\boldsymbol{\pi^\pm \nu_\tau}$ For the same decay set up, *Fig. 9b* features acoplanarity of the planes built respectively on decay products of $\boldsymbol{\tau^+}$ and $\boldsymbol{\tau^-}$. The spin effect is indeed large. However, it requires use of unobservable neutrino momenta. It is difficult or even impossible to achieve sufficient experimental precision in reconstruction of the reaction frame necessary for this observable. In addition, the first observable presented on *Fig. 9a* suffers from the same limitation.

The two other tests, Figures *Fig. 10a* and *Fig. 10b* present distribution of acoplanarity angle for the two planes built respectively on the momenta of $\pi^+\pi^0$ *and* $\pi^-\pi^0$ - the decay products of $\rho^+$ *and* $\rho^-$. All in the rest frame of the ρ-pair. It is directly based on measurable quantities. The $\rho^\pm$ originate respectively from $\tau^\pm \to \nu\, \rho^\pm$ decays. There is no need for Higgs rest frame reconstruction in this case. Events are divided into two categories. If the energy difference between charged and neutral pions coming from the two τ's is of the same sign, they contribute to Fig. *Fig. 10a*, otherwise they contribute to *Fig. 10b*. For details of the definition and for more numerical results obtained with the FORTRAN Interface, see [32].

[11] In the plot the case of non zero scalar-pseudoscalar mixing was chosen. This is the origin of the difference with ref. [31].

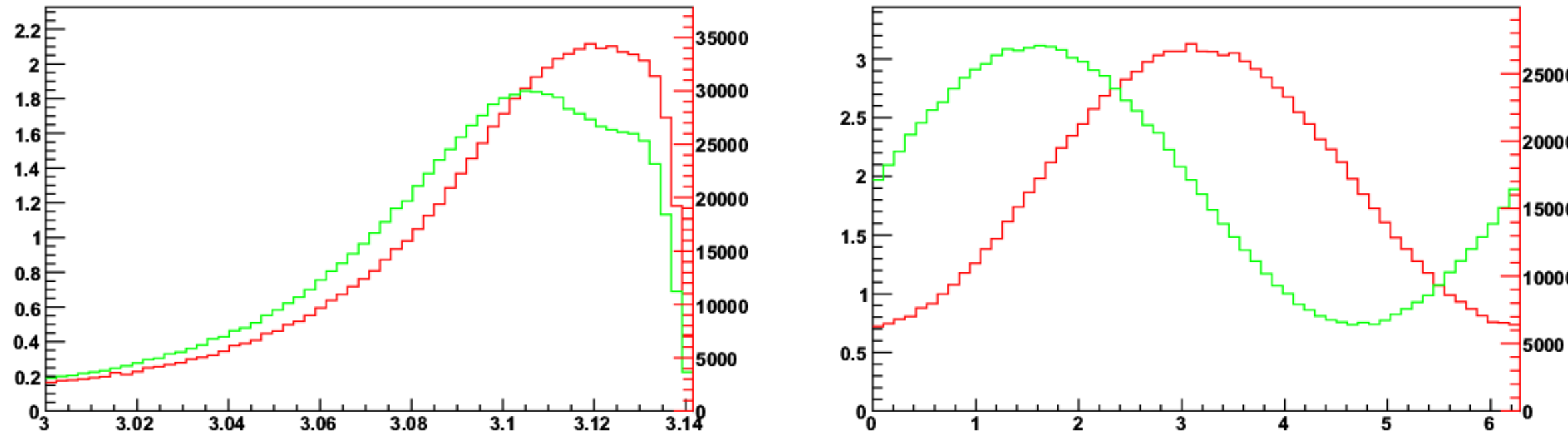

a) $\pi^+ \pi^-$ acollinearity distribution $\sim\pi$ b) $\pi^+ \pi^-$ acoplanarity distribution

*Fig. 9 Transverse spin observables for the Higgs boson for* $\boldsymbol{\tau^\pm \to \pi^\pm \nu_\tau}$*.Distributions are shown for scalar Higgs (red), scalar-pseudoscalar Higgs with mixing angle* $\frac{\pi}{4}$

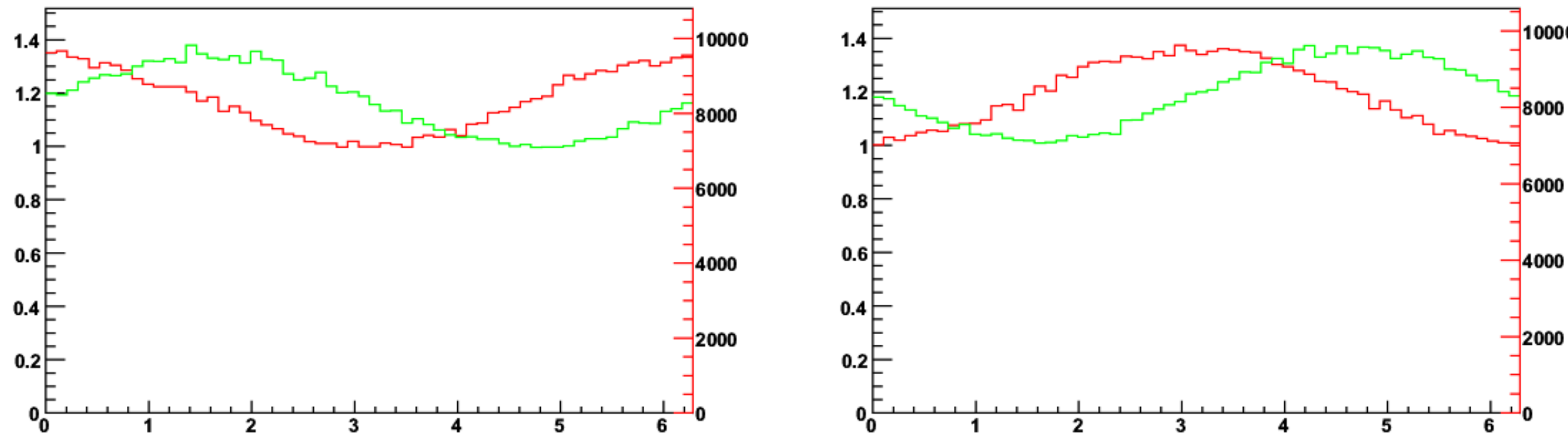

a) acoplanarity distribution $(y_1 y_2 > 0)$ b) acoplanarity distribution $(y_1 y_2 < 0)$

*Fig. 10 Transverse spin observables for the Higgs boson for* $\boldsymbol{\tau^\pm \to \pi^\pm \pi^0 \nu_\tau}$*. Distributions are shown for scalar Higgs (red), scalar-pseudoscalar Higgs with mixing angle* $\frac{\pi}{4}$ *(green)*

## 5.4 $W^\pm \to \tau^\pm \nu_\tau$ and $H^\pm \to \tau^\pm \nu_\tau$

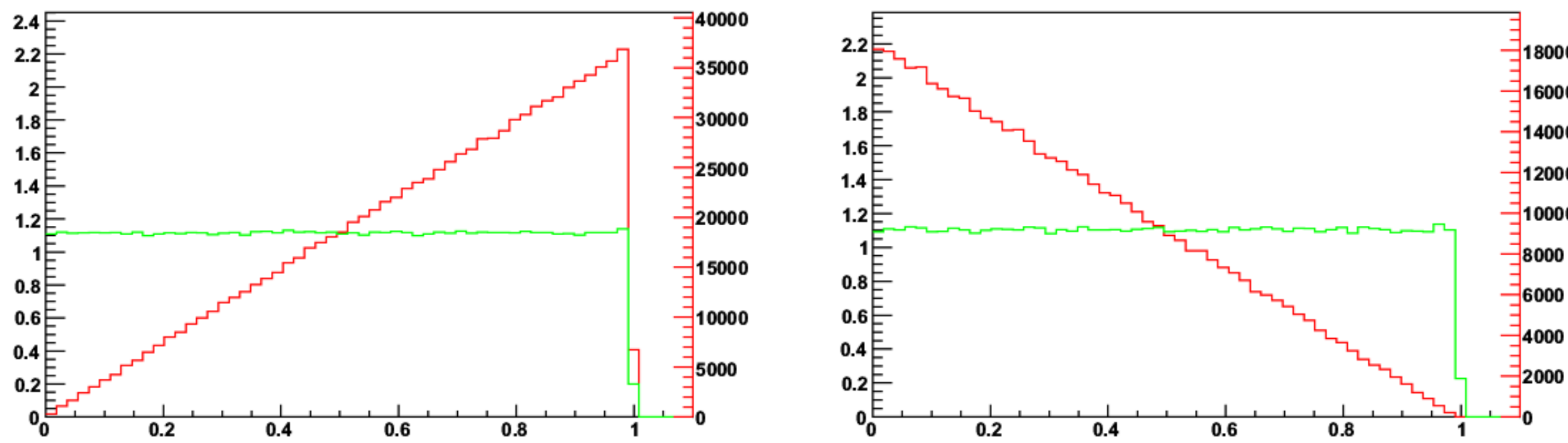

a) $W^+ \to \nu_\tau \tau^+, \quad \tau^+ \to \pi^+ \bar{\nu}_\tau$ b) $H^+ \to \nu_\tau \tau^+, \quad \tau^+ \to \pi^+ \bar{\nu}_\tau$

*Fig. 11 Pion energy spectrum in the rest frame of W (left hand side) and* $\boldsymbol{H^+}$ *(right-hand side). Spin effects included (red line) and neglected (green line) are plotted. The variable* $\boldsymbol{1 - 2\frac{E_{\pi^+}}{M_{W^+}}}$ *or* $\boldsymbol{1 - 2\frac{E_{\pi^+}}{M_{H^+}}}$ *is used respectively*

For the simplest decay mode $\tau^{\pm} \rightarrow \pi^{\pm} \nu_{\tau}$ as was already discussed in ref. [30], the pion energy spectrum should be softer in the case of $W^{\pm}$ decays and harder in the case of charged Higgs decay. This is indeed reproduced in Figs. *Fig. 11a* and *Fig. 11b* and the spectra are reversed for the two cases.

## 5.5 Test of electroweak corrections

One may wonder whether the numerical results induced by electroweak corrections are of any practical purpose. They are expected to be of the order of 1% and indeed are not that large for the intermediate state virtuality of up to 100 GeV above the Z boson mass. The situation changes however significantly at higher energies. As can be seen from Figures *Fig. 12* and *Fig. 13* the effect may be of the order of even 50% at virtualities of several TeV. This is quite in agreement with the results of refs. [33; 34]. In Figures *Fig. 14a* and *Fig. 14b* we collect results for τ polarization calculated at cos(θ)=-0.2. Again, the effects are small up to the energy scale of about 500 GeV. At larger scales corrections become sizable. The electroweak corrections should be therefore considered in studies aiming for new physics phenomena such as $Z' \rightarrow \tau^{+}\tau^{-}$ decays.

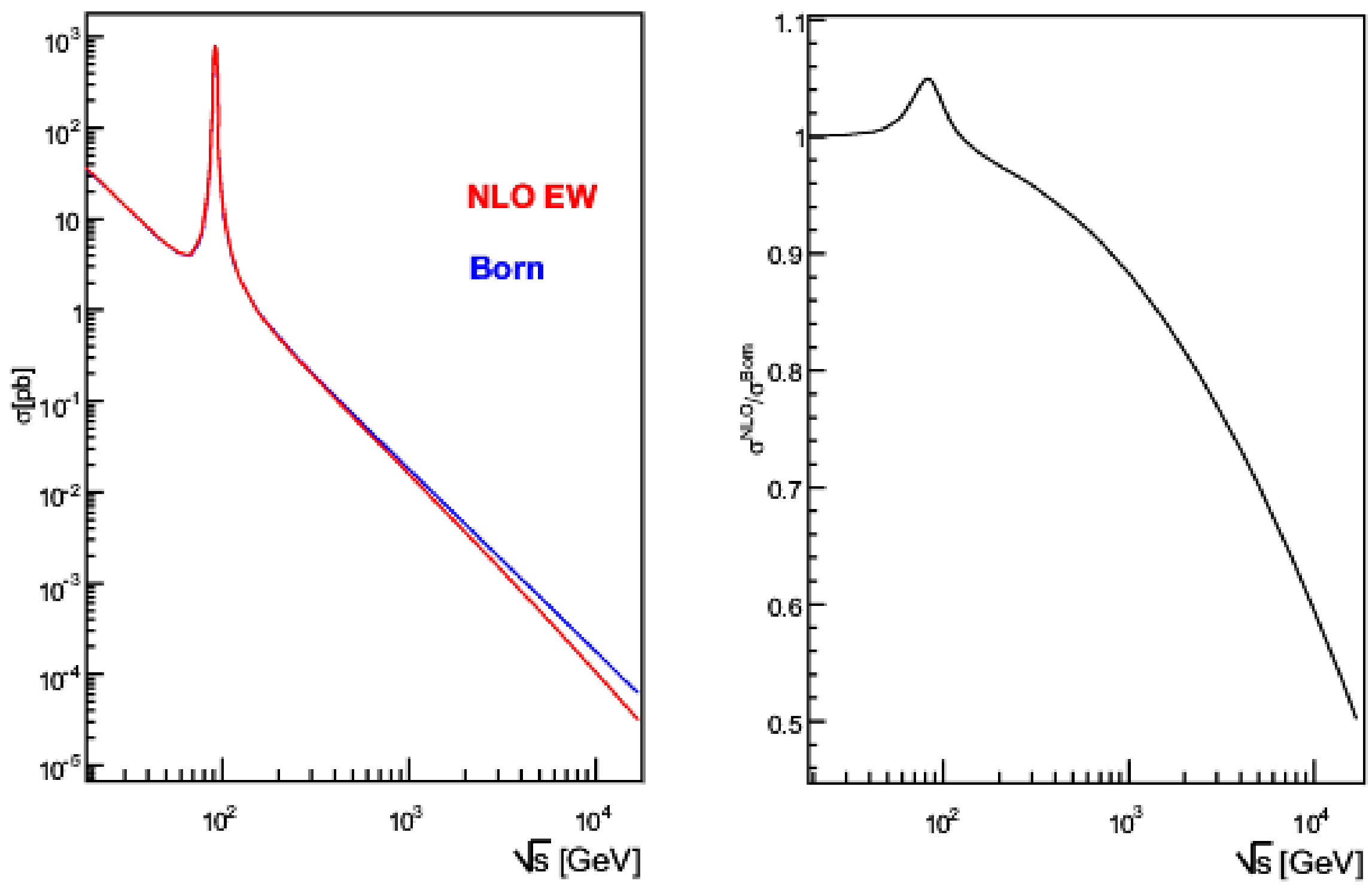

*Fig. 12 The integrated cross section of τ pair production from up quarks calculated with and without NLO EW corrections (red and blue lines) is shown in the left hand side plot. The ratio of the two distributions is given on the right hand plot. We are using the alpha scheme for electroweak corrections. That is why light fermion loops contribute to the difference between the two lines.*

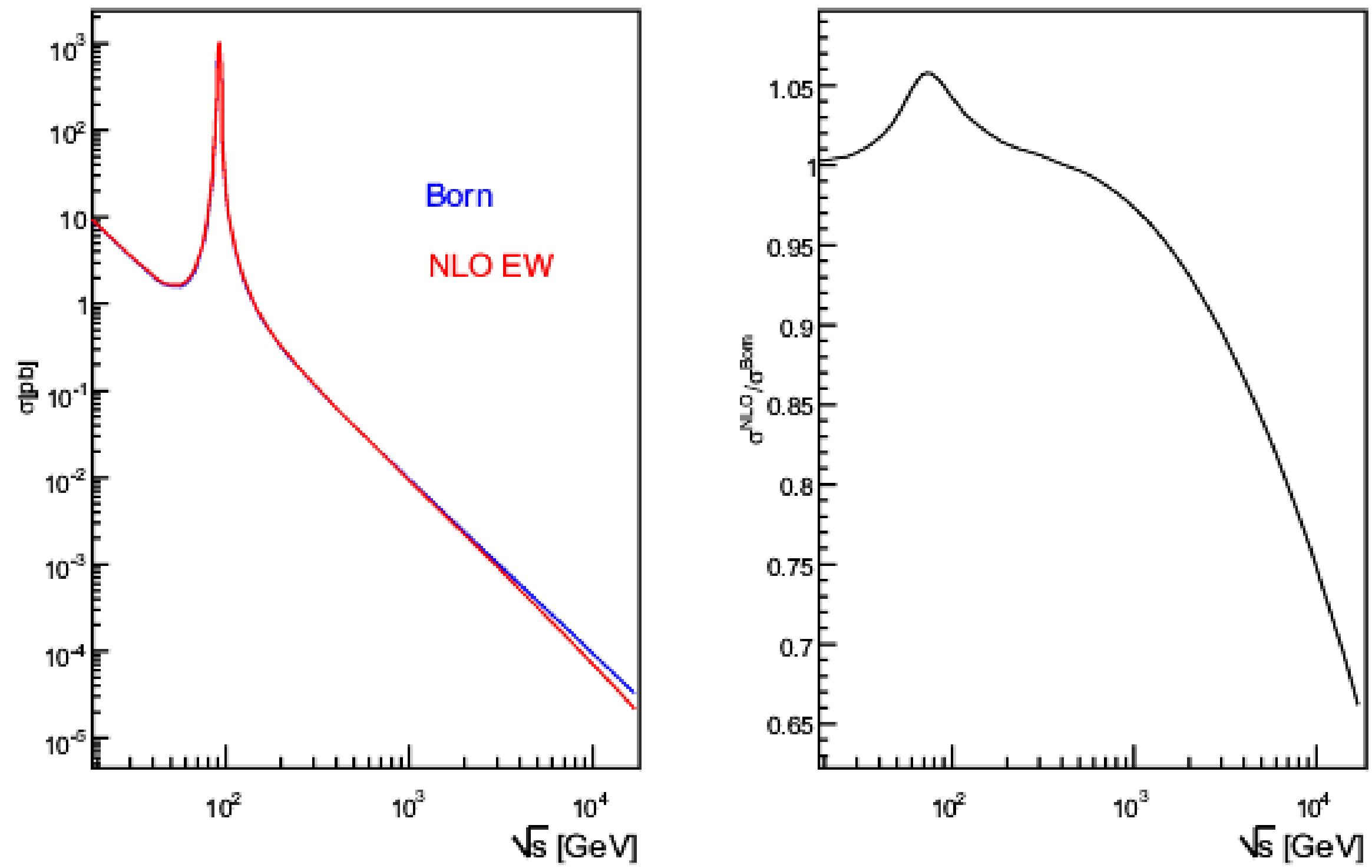

*Fig. 13 The integrated cross section of τ pair production from down quarks calculated with and without NLO EW corrections (red and blue lines) is shown in the left hand side plot. The ratio of the two distributions is given on the right hand plot. We are using the alpha scheme for electroweak corrections. That is why light fermion loops contribute to the difference between the two lines.*

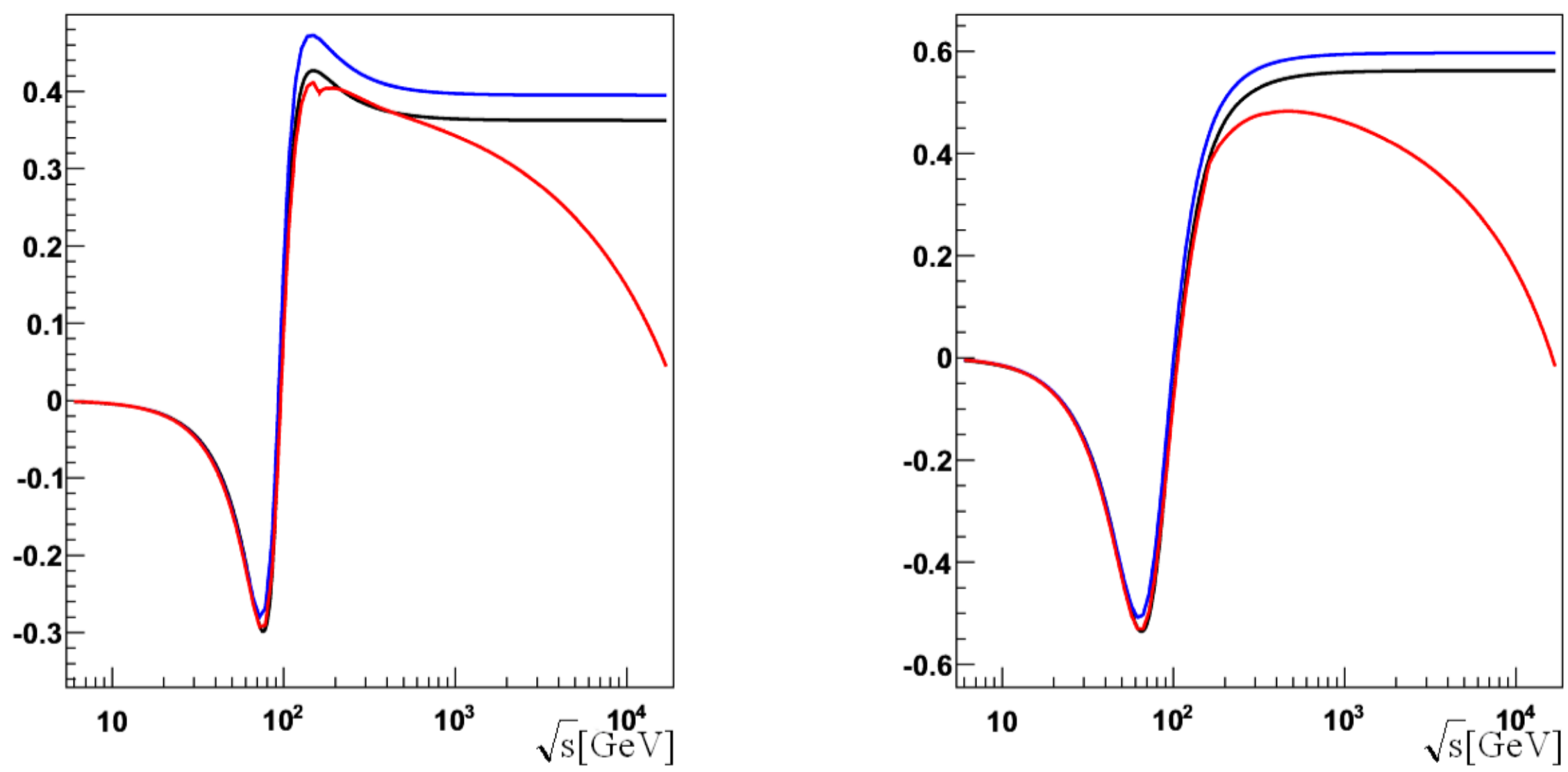

*a)* *b)*

*Fig. 14 Polarization for τ leptons produced from up quarks (Fig. a) and down quarks (Fig. b) at cos(ϑ)=-0.2. The red line is with electroweak corrections, the black is Standard Born as is default in the interface. The blue line is Born according to alpha scheme. The main purpose of these results is a technical test of the software installation. Note however the inadeq7uateness of the alpha scheme Born, which is significantly different from the other two results even at relatively low energies. The small bump on the red line on Fig. (a) is due to the WW threshold. It is insignificant for positive cos(ϑ).*

## 5.6 Testing τ decays

It is important to check, if the τ decays themselves, are generated correctly on the user platform. For this purpose, a separate test routine has been prepared which performs a standard MC−TESTER comparison of program execution with the pre-generated one (of 10 million events). In this case, all τ decay modes are activated and MC−TESTER is simply analyzing τ decays themselves.

| Decay channel | Branching Ratio ± Rough Errors | | Max. shape dif. param. |
|---|---|---|---|
| | Generator #1 | Generator #2 | |
| $\tau^- \to \pi^-\pi^0\nu_\tau$ | 25.3906 ± 0.0504% | 25.3142 ± 0.0503% | 0.02225 |
| $\tau^- \to \nu_\tau\tilde{\nu}_e e^-$ | 17.8393 ± 0.0422% | 18.1628 ± 0.0426% | 0.00000 |
| $\tau^- \to \nu_\tau\tilde{\nu}_\mu\mu^-$ | 17.3381 ± 0.0416% | 17.6346 ± 0.0420% | 0.00000 |
| $\tau^- \to \pi^-\nu_\tau$ | 11.0852 ± 0.0333% | 11.1480 ± 0.0334% | 0.00000 |
| $\tau^- \to \pi^-\pi^0\pi^0\nu_\tau$ | 9.1620 ± 0.0303% | 9.0735 ± 0.0301% | 0.06704 |
| $\tau^- \to \pi^-\pi^-\pi^+\nu_\tau$ | 9.0165 ± 0.0300% | 8.8744 ± 0.0298% | 0.06716 |
| $\tau^- \to \pi^-\pi^-\pi^+\pi^0\nu_\tau$ | 4.3152 ± 0.0208% | 4.5184 ± 0.0213% | 0.26054 |
| $\tau^- \to \pi^-\pi^0\pi^0\pi^0\nu_\tau$ | 1.0912 ± 0.0104% | 0.9995 ± 0.0100% | 0.00000 |
| $\tau^- \to K^-\nu_\tau$ | 0.7187 ± 0.0085% | 0.7184 ± 0.0085% | 0.00000 |
| $\tau^- \to \pi^-\pi^-\pi^+\pi^0\pi^0\nu_\tau$ | 0.5065 ± 0.0071% | 0.0877 ± 0.0030% | 0.00000 |

*Fig. 15 Ten most populous (out of 30) decay channels of* $\tau^-$ *taken from the summary page of MC-TESTER analysis. Statistically significant discrepancies between the benchmark (red) and the newly generated decays (green) can be easily localized and inspected further on plots of the MC-TESTER comparison booklet.*

As seen on *Fig. 15* our test can point out possible problems. In this case the differences in decay channels are due to different FORTRAN TAUOLA version used. The benchmark files were generated using TAUOLA−CLEO while the test files use TAUOLA−ALEPH[12] introducing different form-factor parameterization.

As it was shown in [3] there is plenty of room for improvements in description of τ decays. For this purpose, data from Belle and BaBar should be used. This test as well as similar other tests can be useful for future checks if physics changes are of interest for e.g. LHC phenomenology. In any case, our software is organized to make updates rather straightforward.

# 6. Installation and user interaction

Out of the two projects developed at the same time, the TAUOLA C++ interface and new MC−TESTER version, we have decided that the installation procedure should be performed first on MC−TESTER as it is necessary to test the installation of TAUOLA interface. Whenever a problem with MC−TESTER installation under different environments was found and solved we simultaneously prepared the same patches for TAUOLA so that introducing the interface would go smoothly after all problems with MC−TESTER have been resolved.

[12] Details regarding differences between these versions can be found in reference [16].

## 6.1 ATLAS

Simulations for ATLAS experiment are created using simulation environment called Athena. As described in reference [35]: "*The Athena framework is an enhanced version of the Gaudi framework* [36] *that was originally developed by the LHCb experiment, but is now a common ATLAS-LHCb project and is in use by several other experiments including GLAST and HARP. Athena and Gaudi are concrete realizations of a component-based architecture (also called Gaudi) which was designed for a wide range of physics data-processing applications. The fact that it is component-based has allowed flexibility in developing both a range of shared components and, where appropriate, components that are specific to the particular experiment and better meet its particular requirements.*"

Athena uses unique procedure to register generators in a queue distributed to the computer grid. Each generator need its own interface designed for this framework. Fortunately, older version of MC-TESTER and TAUOLA had such interface prepared so we already had a strong starting point to create new interfaces. However, it is worth to note that such module, as well as few other scripts described here, had to be done independently from the rest of the code as a means to install the software into Athena environment. This kind of specialized code cannot be used for any other purpose.

Despite the information that HepMC 2.05 has been introduced as the new standard for experiments, at the moment of installation, and to this day, Athena operated on the older version, 2.03. As described in HepMC release notes [23], the transition between versions 2.03 and 2.04 was quite major as it introduced the units system used to adapt to different momentum and length units used in experiments. Since our software was specifically converted during development to use this system, in order for our software to work under Athena we had to downgrade it to HepMC version 2.03.

Of course such downgrade had to be done only to Athena environment and only temporary, as sooner or later HepMC 2.05 would have to be installed on Athena as well, therefore we have created it as a means of a shell script that user can execute independently from the installation. The script was designed to modify the code if the user platform had HepMC version lower than 2.04. To create a user-friendly solution, the script was incorporated into the configuration procedure so that depending on the HepMC version located by autoconf script, the configuration could downgrade our software automatically.

Unfortunately, that was not the only problem we have run into. After several trials, thanks to Marcin Wolter, Anna Kaczmarska and Eric Torrence, we have figured out that Athena was working with an older, FORTRAN version of PYTHIA Monte Carlo generator – PYTHIA 6. The problem with this generator was that it already used its own version of TAUOLA FORTRAN – similar to the one used in our software as an independent module. There would be no problem with that if it were not for the collision in function names [37].

Since FORTRAN names its symbols located in the library in a static way that cannot be changed, the solution that allowed us to communicate between FORTRAN and C++ code by overwriting the routine filhep_ routine, failed as Athena took precedence in symbols loaded from internal libraries rather than those from a generator attached externally. Even though decays were generated properly, they could not be saved into HepMC event record.

This was an unexpected result and we could not anticipate such problem designing our solution. Yet again, it was a problem unique to Athena environment and also seemed to be tem-porary as there were already plans to switch the Athena Monte Carlo generator to `PYTHIA 8.1`. The tests with this generator under Athena worked without any problems, but in order to allow testing the new C++ interface until PYTHIA 8.1 is introduced, which could take anywhere from a month to a year, we had to come up with yet another fix specific for ATLAS experiment and independent from the rest of the project.

Since the ATLAS was the first platform we have started tests on, we have figured out that similar problems will occur with other experiments as well. Even though collaboration tries to introduce a single universal standard, separate experiments still focus on their own software management systems and are reluctant to change it as it has proven to work correctly. Therefore, each new standard takes different amount of time to be introduced in different experiments. Fortunately, the problem was not hard to resolve and was fixed by renaming all FORTRAN symbols, as well as the overwritten `filhep_` routine, which was done, again, as a shell script that can be executed by the individual user.

## *6.2* CMS

The CMS experiment uses its own software framework called CMSSW. As stated in online sources [38]: "*CMSSW is a large-scale software system for data acquisition, triggering, reconstruction and analysis of CMS data. It introduces Event Data Model based on the single event contained in memory with modular content processed by separate units. Each unit has a well-defined event processing functionality and communication between different steps of the analysis routine is performed only using the event data structure.*" Within CMSSW, the independent projects are managed using Source Configuration, Release, And Management (SCRAM) tool. In order to work under CMS we had to learn this environment and compile our project within it.

The main problem was that none of us had any experience with this environment at all. We had to search for someone within the CMS experiment willing to help us with the installation. We were able to do so thanks to Yulia Yarba and Sami Lehti.

SCRAM uses very strict compilation procedure forcing us to modify our software to meet these standards. Since we are dealing with an old project, parts of the code had to be cleaned up to meet the compiler standards for CMS. In addition, SCRAM had troubles processing dictionaries generation performed by `ROOT` framework. It was a well-known problem and the only solution we were able to achieve was to generate them manually and include them into the installation used by CMS. This operation has to be performed each time the data structure changes, but it does not pose a huge problem, as there should be no need of frequent changes to software architecture.

As with ATLAS, we expected that it would not be the end of our troubles. CMS environment uses its own system of parallel computation that produces a separate output files meant to be merged together after simulation ends. Unfortunately, the default software used for merging the `ROOT` output files failed in case of `TAUOLA` result files. Consequently, a separate `ROOT` script had to be written that allowed merging of these files. Such functionality was not needed before, but it might become useful in other cases so we decided to add it to main software distribution.

Finally, the long analysis procedure using several input files proved to create a memory management problems that, to this day, remain unresolved. Hopefully, with better under-standing of CMS environment we will be able to figure out and solve the problem as soon as possible.

## 6.3 LCG database

Quoting the introduction from LCG webpage [39]: "*LHC Computing Grid (LCG) is a global collaboration producing a massive distributed computing infrastructure that provides more than eight thousands physicists around the world with near real-time access to LHC data, and the power to process it. LCG Project manages a multi-platform universal database of all projects available for all experiments around LHC. It is designed to build and maintain data storage and analysis infrastructure for the entire high energy physics community that will use the LHC.*" Software installation under LCG was a necessary step in order to distribute our software to all interested users with unified installation procedure for all platforms.

Since the software must be designed to work on all machines connected to a grid, it has to be compatible with different variants of Scientific Linux [40] installed at CERN and different OS/X installations along with several compiler versions for both 32bit and 64bit architecture. For this purpose, a series of autotools scripts has been created and adopted by Oleg Zenin and Dmitri Konstantinov.

This method supposed to work under any platform, however, as with all software designed to be universal, there are always a compatibility issues. In case of scripts generated by autotools, the problem regards the `ROOT` configuration under OS/X and its library `freetype`, `which` does not exist under OS/X installations. It is a common problem and even though this library is not used in our project, the appropriate line of configuration script containing this directive had to be modified by hand. This proves once more that even the smallest things may break the automated processes creating a need for individual solutions.

In the end, scripts used for LCG installation are very hard to modify manually and are nearly useless to the regular user. In case of unique platform on which user would like to install the software, they provide very limited help on how to solve the installation issues. Moreover, any future modification of these scripts would be hard to implement as the base scripts are created for LCG purpose only. That is why we decided to stick to our previous configuration option based only on autoconf software which we know exactly how to modify and can apply to any destination platform required.

Therefore, the LCG scripts had been used to install the software into the database, and had been included in our software as an alternative means of installation that can be activated using shell script. This way users familiar with LCG installation procedure can use these scripts if they want to, while everyone else may use our configuration procedure or modify the platform dependant options by themselves, which introduces even more flexibility to the end user.

## 6.4 User feedback

As soon as project became accessible to the users, we have received a lot of feedback, which allowed us to customize our software even further. This experience expressed yet another difference between the commercial software and project created for wide collaboration. Since there is no single customer who would show his list of expectations and pointed out which parts of the code needs to be improved and which functionality should be implemented, there is a wide variety of users that can express their opinion and request limitless number of options that were impossible to predict at the planning step. Here I present the few implemented functionalities requested by the users. They serve as an example of how unique expectations users may have.

- Random Generator substitution
  The core of Monte Carlo process is the random generator used within the algorithm. Since some experiments require strict verification and full control of the generator itself, we wrote an option `Tauola::setRandomGenerator(...)` allowing the user to substitute built-in random number generator with the one provided by the user.
- Single Tau decay
  There are cases in which the whole event structure is not needed for the particular analysis of a branch of decay tree. In such cases user might prefer to decay a single τ manually inside his own analysis routine. In such cases, the polarization information has to be provided by hand. `Tauola::decayOne(...)` provides such option generating single τ decay with or without the polarization information.
- External boosting routine
  An advanced analysis might require that the decayed τ along with its daughters should be boosted to a custom laboratory frame. Combined with the above option, `Tauola::setBoostRoutine(...)` gives flexibility that allows creating a wide range of specialized tests.

## 6.5 Future plans

The development and installation of MC-TESTER is finished. Next step is further extending TAUOLA functionality according to new user requests and adopting both projects according to changes within the collaboration.

Apart from MC-TESTER and TAUOLA C++ interface, a third project needs to follow the same routine. The abstract model of PHOTOS C++ interface [5] has already been created and the first functionality has been implemented, but the project is far from being completed. Similar to TAUOLA, PHOTOS is a large project requiring specialized approach and extensive knowledge in physics. However, it is well defined and with the basic experience gained from TAUOLA, the first steps of development of this project become much easier.

Lastly, an extension to low energy project TAUOLA FORTRAN is planned. It regards post-generation filtering of decay models [3]. In future, the description of hadronic currents might change, a dynamic methodology to introduce and test such changes will be needed. Such currents are visible through a set of static distributions. The idea is to create an external environment that allows modifying these distributions in order to fit them to already generated data sets. Thanks to this, there will be no need to generate new set of events to test a particular model, saving considerable amount of time as data generation can take hours or days for large data sets.

As physics and software engineering blends together, the software development becomes much harder as both coding and testing requires extensive research and a large amount of time on validation of the results. Thanks to the experience gained during development of this project, I hope to avoid repeating the same mistakes and exert this knowledge to overcome any new problems.

## 7. Summary

The development process described here does not include development of abstract level of methods for program tests. Such methods were already created for previous version of the project and were sufficient for our needs. That is why this challenging aspect of large project development requiring cooperation between programmers and physicists is not discussed here.

Starting from basic model prepared by Nadia Davidson and Zbigniew Wąs, my part of the contribution to the project consists of:

- Development of user configuration module, in particular writing options described in sections 4.5.3 to 4.5.6 as well as options created thanks to user feedback (see section 6.4 for few examples of such options),
- Creation of configuration scripts and adjusting the installation procedure to user needs,
- Introduction of Electroweak Corrections module created by Gizo Nanava,
- Debugging, memory leak tracking and testing.

Nearing competition of first stable release, I assisted Zbigniew Wąs with installation of the software within the collaboration experiments (described in Section 5.6). Last, but not least, my responsibility was to assure that program remain coherent and to create throughout tests of consecutive versions.

I have started on the project related to my thesis in fall 2008. The first stable version of `TAUOLA` has been released on 25 January 2010. As software gains more popularity, more user feedback becomes available and new functionality has to be implemented, which show that integration of this software within the collaboration has been successful. All what is left is to maintain the software introducing new functionality and implementing new physics processes currently under test. The pre-print version of the documentation is available under arXiv database [1]. Details regarding the stable `TAUOLA` release and the newest features, as well as doxygen documentation, can be found on `TAUOLA` webpage `http://www.ph.unimelb.edu.au/~ndavidson/tauola/doxygen/`.

The results of work on `MC-TESTER` has been published as a pre-print under arXiv database [2] and its final version containing the update mentioned here has already been prepared. `MC-TESTER 1.24` has been successfully installed under ATLAS environment and we are currently performing last step of installation under CMS. It has also been installed within the LHC computing grid; users with access to LCG can find `MC-TESTER` on `/afs/cern.ch/sw/lcg/external/MCGenerators/mctester` directory. More information, as well as download of the newest version of `MC-TESTER`, can be found on the webpage `http://mc-tester.web.cern.ch/MC-TESTER/`.

The experience gained from this project helped me understand two aspects of the abstract model implementation. The first being the creation of the large structure defining the behavior of each component of the project including structure-related algorithms and the main algorithm outline. The second is the numerical programming of formulas, defining and implementing solution for specific exceptions from the regular algorithm behavior and resolving numerical stability issues. The first one can be described as looking on the software from a macro scale. Typically, the difficulty is to design an appropriate abstract model and to maintain its changes, which in case of high energy physics project, can occur at any time. The second, which in comparison can be called microprogramming, on the first glance seems to be easy, showing its true meaning during the most complex tests. Especially tests focused on the precision.

I have also used this project as opportunity to understand relation of physics-driven software development as well as practical case concepts of agile-based programming methodologies. Both similarities and differences presented in Section 4.1 expanded my knowledge about practical implementation of the theory behind the agile approach. This aspect of my work is of course not conclusive. I am looking forward to progress on it in my professional life.